\documentclass{aa}
\usepackage{natbib}
\usepackage[varg]{txfonts}
\usepackage{hyperref}
\usepackage[dvipsnames]{xcolor}
\hypersetup{
    colorlinks,
    linkcolor={blue!90!black},
    citecolor={blue!90!black},
    urlcolor={blue}
}
\usepackage[T1]{fontenc}
\usepackage{ae,aecompl}
\usepackage{graphicx}
\usepackage{amsmath}
\usepackage{float}
\usepackage{amssymb}
\usepackage{multirow,bigdelim}
\usepackage{mathtools}
\usepackage{siunitx}
\usepackage{tabulary}
\usepackage{tablefootnote}
\usepackage{threeparttable}

\newcommand{\tr}{}

\newcommand{\sqdeg}{\,{\rm{deg}}^{2}}
\newcommand{\wth}{$w(\vartheta)$ }
\newcommand{\nflask}{30 }
\newcommand{\flask}{{\sc{flask}} }

\newcommand{\eg}{e.g. }
\newcommand{\ie}{i.e. }
\newcommand{\ttt}[1]{\texttt{#1}}

\newcommand{\tbf}[1]{{#1}}
\newcommand{\tbi}[1]{{#1}}
\newcommand{\hlstop}[1]{{#1}}

\usepackage{bm}
\usepackage{soul}
\usepackage{comment}
\numberwithin{equation}{section}

\begin{document}

\title{Organised randoms: Learning and correcting for systematic galaxy clustering patterns in KiDS using self-organising maps}
\titlerunning{Organised randoms}

\author{
Harry Johnston\inst{1,2}\thanks{h.s.johnston@uu.nl},
Angus H. Wright\inst{3},
Benjamin Joachimi\inst{2},
Maciej Bilicki\inst{4},
Nora Elisa Chisari\inst{1},
Andrej Dvornik\inst{3},
Thomas Erben\inst{5},
Benjamin Giblin\inst{6},
Catherine Heymans\inst{3,6},
Hendrik Hildebrandt\inst{3},
Henk Hoekstra\inst{7},
Shahab Joudaki\inst{8,9},
Mohammadjavad Vakili\inst{7},
}

\authorrunning{H. Johnston, A. H. Wright et al.}

\institute{
Institute for Theoretical Physics, Utrecht University, Princetonplein 5, 3584 CE Utrecht, The Netherlands\and
Department of Physics and Astronomy, University College London, Gower Street, London WC1E 6BT, UK\and
Ruhr University Bochum, Faculty of Physics and Astronomy, Astronomical Institute (AIRUB), German Centre for Cosmological Lensing, 44780 Bochum, Germany\and
Center for Theoretical Physics, Polish Academy of Sciences, al. Lotnik\'{o}w 32/46, 02-668 Warsaw, Poland\and
Argelander-Institut f\"{u}r Astronomie, Auf dem H\"{u}gel 71, 53121 Bonn, Germany\and
Institute for Astronomy, University of Edinburgh, Royal Observatory, Blackford Hill, Edinburgh, EH9 3HJ, UK\and
Leiden Observatory, Leiden University, PO Box 9513, Leiden, NL-2300 RA, the Netherlands\and
Department of Physics, University of Oxford, Keble Road, Oxford, OX1 3RH, UK\and
Waterloo Centre for Astrophysics, University of Waterloo, 200 University Ave W, Waterloo, ON N2L 3G1, Canada}
\date{Accepted XXX. Received YYY; in original form ZZZ}


\abstract{
We present a new method for the mitigation of observational systematic effects in angular galaxy clustering through \hlstop{the use of} corrective random galaxy catalogues. Real and synthetic galaxy data from the Kilo Degree Survey's (KiDS) 4th Data Release (KiDS-$1000$) and the Full-sky Lognormal Astro-fields Simulation Kit ({\sc{flask}}) package, respectively, are used to train self-organising maps to learn the multivariate relationships between observed galaxy number density and up to six systematic-tracer variables, including seeing, Galactic dust extinction, and Galactic stellar density. We then create `organised' randoms\hlstop{;} random galaxy catalogues with spatially variable number densities, mimicking the learnt systematic density modes in the data. Using realistically biased mock data, we show that these organised randoms consistently subtract spurious density modes from the two-point angular correlation function $w(\vartheta)$, correcting biases of up to $12\sigma$ in the mean clustering amplitude to as low as $0.1\sigma$, over an  angular range of $7-100\,\rm{arcmin}$ with high signal-to-noise ratio. Their performance is also validated for angular clustering cross-correlations in a bright, flux-limited subset of KiDS-$1000$, comparing against an analogous sample constructed from highly complete spectroscopic redshift data. Each organised random catalogue object is a clone carrying the properties of a real galaxy, and is distributed throughout the survey footprint according to the position \hlstop{of the parent galaxy} in systematics space. Thus, sub-sample randoms are readily derived from a single master random catalogue through the same selection as applied to the real galaxies. Our method is expected to improve in performance with increased survey area, galaxy number density, and systematic contamination, making organised randoms extremely promising for current and future clustering analyses of faint samples.
}

\maketitle

\section{Introduction}
\label{kids:sec:intro}

Recent decades have seen the advent of precision cosmology, as inferred from the large-scale structures of the Universe \citep{Ho2012,Font-Ribera2014,Anderson2014,Hildebrandt2016,Alam2017,Troxel2017,Hamana20,eBOSS2020,Troster2020,Asgari2020b}. We can shed light on the mysterious dark matter scaffold by examining the distribution of the galaxies supported by it, and extend these measurements across broad redshift epochs in order to study dark energy through the evolution of structures and the universal expansion history. \hlstop{Tensions are beginning to emerge between near-universe} \citep{Freedman2017,Riess2019,Hildebrandt2018,Joudaki2019,Heymans2020} \hlstop{and cosmic microwave background} \citep[CMB, e.g.][]{PlanckCollaboration2018} \hlstop{measures of the expansion rate, and the amount and clustering of matter in the Universe.} Resolving these \hlstop{tensions} will require strict control of sources of systematic error in the new era of statistical precision soon to be explored (Dark Energy Spectroscopic Instrument; \citealt{Aguilar}, Rubin Observatory; \citealt{LSSTScienceCollaboration2009}, \emph{Euclid}; \citealt{Laureijs2011}).

Currently, the most competitive large-scale structure (LSS) constraints upon the matter energy density $\Omega_{\rm{m}}$ versus normalisation of the matter power spectrum $\sigma_{8}$ plane come from combined analyses of galaxy clustering and weak gravitational lensing \citep{VanUitert2018,Joudaki2017,DESCollaboration2017,Asgari2020a,Heymans2020}, making use of wide-area galaxy survey data (\eg Dark Energy Survey; \citealt{DES2005}, Kilo Degree Survey; \citealt{DeJong2013}, Hyper Suprime-Cam; \citealt{Miyazaki2012}). However, these surveys are susceptible to complex source selection functions, which can destroy and/or mimic real, informative signals, such as cosmological density fluctuations. These selection functions are generally imprinted on galaxy survey data in a way that correlates with observational and physical properties of the survey on-sky. Examples of such properties, which have the propensity to introduce spurious selection effects on galaxy data, include spatial variation in atmospheric seeing, the telescope point-spread function (PSF), stellar density, Galactic dust extinction, and more. Crucially, it is possible, or even likely, that some spurious selection functions may be introduced due to a complex confluence of many observational properties, some of which we may be unaware of.

Upcoming surveys are likely to build upon the many \hlstop{weak lensing} analyses that currently aim to constrain cosmology with the 3x2pt analysis method: simultaneously fitting the two-point functions describing the galaxy shear-shear (cosmic shear), position-shear (galaxy-galaxy lensing), and position-position (galaxy clustering) correlations. Analysed simultaneously, these statistics help to constrain nuisance parameters (\eg photometric redshift calibration uncertainties, or intrinsic alignments), thereby tightening cosmological constraints. However, this additional constraining power also brings sensitivity to systematic biases; any bias in (e.g.) the galaxy clustering correlation, such as those that may be introduced by systematic variation of the galaxy density field, will propagate into the inferred (joint) cosmology in a pathological fashion.

\tbi{This work focuses on sources of systematic error affecting the detection of galaxies, and propagating into statistics involving galaxy positional data, for instance correlation functions. Following recent studies on this topic, it is useful to divide mitigation strategies into three categories: (i) Monte Carlo simulation of synthetic objects, (ii) mode (de)projection, and (iii) regression.}

\tbi{The first approach seeks to inject artificial galaxies into realistic images, so as to gauge the detector-response to observational systematics \citep{Berge2013,Suchyta2016}. As a result, a forward model for the survey selection function can be built, from which a Monte Carlo sampling can generate mock catalogues of unprecedented fidelity. This computationally demanding technique is extremely promising and under active development \citep{Everett2020}.}

\tbi{Techniques for systematics mode projection, for example assigning large variance to spurious modes so that they are ignored by power spectrum estimators, were developed by \cite{Leistedt2013} and \cite{Leistedt2014} for photometric quasar clustering, and similar methods have recently been applied to Hyper Suprime-Cam (HSC) data by \cite{Nicola2020}. Interestingly, \cite{Weaverdyck2020} demonstrated that mode projection techniques can be consistently expressed under a regression framework, and showed that extensions to previous formulations can automate the selection of important systematic features and remain robust under scenarios of numerous correlated systematics.}

\tbi{Much work on this topic has focused upon regression-based approaches, where the aim is to model the functional relationships between systematic-tracing variables and galaxy number densities. \cite{Ross2011} and \cite{Ho2012} suppressed systematics in SDSS BOSS-like (Sloan Digital Sky Survey; \citealt{York2000}, Baryon Oscillation Spectroscopic Survey; \citealt{Eisenstein2001}) photometric luminous red galaxy (LRG) data by deriving per-galaxy inverse weights from number density-systematics (‘one-point’ or ‘pixel’) correlations or by computing signal corrections (assuming systematic-tracers relate linearly to galaxy number densities) from galaxy-systematic (two-point) cross-correlations. \cite{Vakili2020} also estimated weights from one-point functions to measure the clustering of photometric LRGs in the Kilo Degree Survey (KiDS), decomposing systematic-tracers in an orthogonal basis and exploring second-order polynomials to characterise cross-talk between parameters. \cite{Elvin-Poole2017} iteratively derived weights for Dark Energy Survey (DES) LRGs from linear systematic-density fits, and \cite{Wagoner2020} recently improved upon this analysis by performing simultaneous likelihood fitting of linear coefficients to all systematics maps and the observed density contrast, and then calibrating for over-correction of clustering correlations with mock catalogues.}

\tbi{A possible sub-category of regression is comprised of studies that use machine learning to predict the relationship between observed galaxy number densities and multivariate systematics vectors, and then derive density field corrections without the least-squares methods that typically characterise regression-based approaches. For example, \cite{Rezaie2019} derived weights for emission line galaxies (ELGs) selected (following eBOSS; \citealt{Raichoor2017}) from the Dark Energy Camera Legacy Survey \citep[DECaLS;][]{Dey2019} using deep neural networks. \cite{Morrison2015} tackled clustering biases in the Canada-France-Hawaii Telescope Lensing Survey \citep[CFHTLenS;][]{Erben2013}, using k-means clustering in the high-dimensional density-systematics space to create weight-maps from which to draw random points. These random points then compensate for systematic density fluctuations through typical estimators for clustering correlations. Our methods belong to this machine-learning regression category, which has the advantage over most regression-based methods that the multivariate density-systematics models require no functional form and can be arbitrarily non-linear, allowing for the compensation of more complex modifications to the cosmic density field.}

We seek to mitigate density field biases through the construction of tailored random galaxy catalogues (hereafter `organised randoms'), which mirror systematically induced galaxy-density variations  \citep[similarly to][]{Morrison2015,Suchyta2016}. Random galaxy catalogues (commonly referred to simply as `randoms') are widely used when estimating galaxy clustering and galaxy-galaxy lensing (GGL), whereby correlations between galaxies and random points allow for reductions in methodological bias, improved covariance properties between the statistics, removal of systematics caused by edge or masking effects, and aid in the reduction of additional systematic correlations \citep{Landy1993,Singh2016}. For this task, studies typically employ high-density (relative to the survey galaxy number density) spatially uniform random points. However, while these will aid in the removal of systematic correlations with the observed galaxy distribution, they cannot account for spatial correlations stemming from systematically \emph{\textup{unobserved}} galaxies, or those systematically lost due to sample selection effects.

Our work aims to tackle this problem: we use a form of machine-learning assisted dimensionality reduction, the self-organising map \citep[or `SOM',][]{Kohonen1990}, to infer from the observed galaxy distribution the high-dimensional mapping between survey systematics and galaxy number densities on-sky. We then create many clones of the real galaxies  \citep[\ie copies retaining all photometric or other properties, see][]{Farrow2015} and distribute them as random points throughout the survey footprint, in accordance with their systematically derived number density on-sky. This allows any selection effects in galaxy data to be trivially mirrored in the organised randoms, thereby preserving the systematic patterns and the systematically induced density variations for an arbitrarily defined galaxy sample. 

The paper is organised as follows: In Sec. \ref{kids:sec:data} we introduce our galaxy data from Kilo Degree Survey \citep[KiDS,][]{Kuijken2019} observations and from the Full-sky Lognormal Astro-fields Simulation Kit \citep[{\sc{flask}};][]{Xavier2016} simulations. Sec. \ref{kids:sec:som} describes self-organising maps and how they are used in this work. \tbf{In Sec. \ref{kids:sec:artificial_systematics} we assess the capability of the self-organising map to identify artificially created systematic trends in galaxy density. Sec. \ref{kids:sec:datadriven_systematics} then turns to KiDS data-driven systematic density fluctuations and demonstrates the utility of organised randoms in recovering unbiased clustering signals from realistically biased \flask mocks. Final data applications are presented in Sec. \ref{kids:sec:kids_clustering}, and we make concluding remarks in Sec. \ref{kids:sec:discussion}. Throughout this work, we quote AB magnitudes and assume a fiducial  WMAP9+BAO+SN cosmology \citep{Hinshaw2013}: flat $\Lambda$CDM, with $\Omega_{\rm{m}}=0.2905,\,\Omega_{\rm{b}}=0.0473,\,\sigma_{8}=0.826,\,h=0.6898,\text{and}\,n_{\rm{s}}=0.969$.}

\section{Data}
\label{kids:sec:data}

We validated our random catalogues using both real and synthetic galaxy data, and invoking both realistic distributions of systematic parameters on-sky  \citep[drawn from the KiDS 4th$^{\rm }$ Data Release, DR4,][]{Kuijken2019} and artificially constructed systematics distributions. For our simulations, we used lognormal galaxy fields simulated with \flask \citep{Xavier2016}.

\subsection{KiDS}
\label{kids:sec:kids}

\begin{figure*}[!htp]
    \centering
    \includegraphics[width=\linewidth]{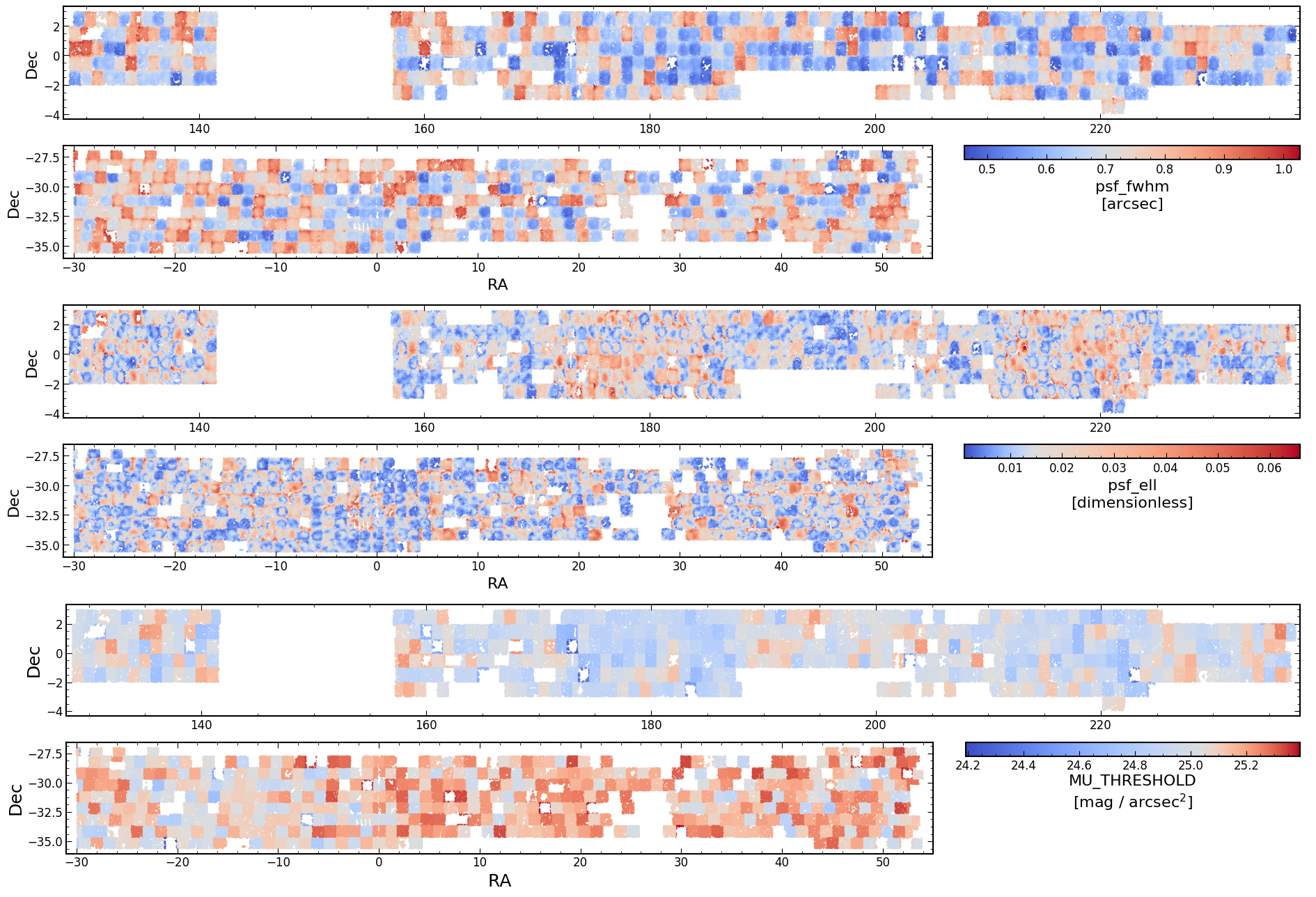}
    \caption[KiDS: maps of systematic-tracer variables in the KiDS bright sample]{Maps of systematic-tracer variables (from Table \ref{kids:tab:housekeeping_variables}) from $r$-band (the detection band) imaging in the KiDS-North (\emph{top panels}) and KiDS-South (\emph{bottom panels}) areas. Grey denotes the $50th^{\rm{}}$ percentile of the systematics distribution in each case, and blue and red then denote good and bad observing conditions relative to the $50th^{\rm{}}$ percentile. As we show in Fig. \ref{kids:fig:SOMfig}, the majority of spatial variations in galaxy number density correlate with these parameters at $\lesssim5\%$.
    }
    \label{kids:fig:Apars}
\end{figure*}

\begin{figure*}[!htp]
    \centering
    \includegraphics[width=\linewidth]{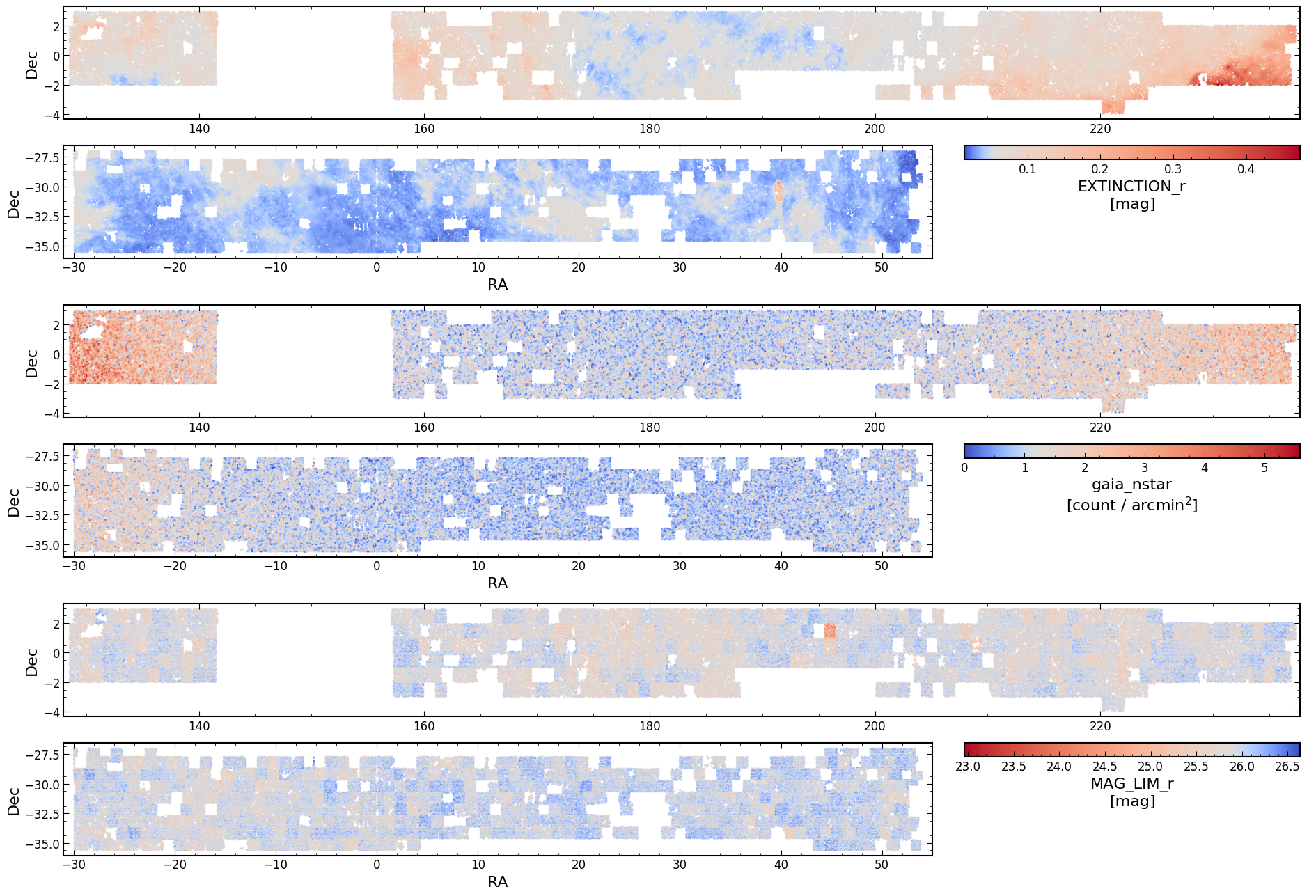}
    \caption[KiDS: more maps of systematic-tracer variables in the KiDS bright sample]{Same as Fig. \ref{kids:fig:Apars}, but for the remaining systematic-tracer variables from Table \ref{kids:tab:housekeeping_variables}.
    }
    \label{kids:fig:BCpars}
\end{figure*}

Both KiDS \citep{DeJong2013} and its partner survey, the VISTA Kilo-Degree Infrared Galaxy (VIKING; \citealt{Edge2013}) survey, are now observationally complete, covering a combined area of $1350\sqdeg$ on-sky in a total of nine photometric bandpasses  ($ugriZY\!JHK_{\rm{s}}$). Over $1000\sqdeg$ of this combined dataset is publically available as part of KiDS DR4\footnote{\url{http://kids.strw.leidenuniv.nl/DR4/}} \citep{Kuijken2019}, providing gravitational shear estimates (unused in this work), nine-band photometric redshift estimates \citep{Hildebrandt2018,Wright2020,Hildebrandt2020}, and observational information for over $100$ million galaxies. This galaxy sample, typically referred to as KiDS-$1000$ (though the retained area after masking is $\sim900\sqdeg$), samples a wide range of observing conditions over a large area, and is consequently imprinted with an unknown combination of systematic galaxy depletion and enhancement patterns. Of all the various data products provided within the KiDS DR4, we identify a selection of systematic-tracer variables (detailed in Table \ref{kids:tab:housekeeping_variables} and mapped out in Figs. \ref{kids:fig:Apars} and \ref{kids:fig:BCpars}) that trace physical phenomena that have the greatest potential to imprint subtle galaxy selection functions on the dataset, based on our experience. These parameters, individually and in combination, form the dataset used to train our SOMs. It is not known a priori whether the chosen variables all trace real observational phenomena that cause the systematic loss of galaxies; we therefore explore the effect of unimportant, `distracting' variables as we test our methodology (Sec. \ref{kids:sec:artificial_systematics_design}).

\begin{table*}[!htbp]
\centering
\begin{threeparttable}
    \def\arraystretch{1.1}
    \caption[KiDS: systematic-tracer variables for training of self-organising maps, with units and descriptions]
    {KiDS: systematic-tracer variables for training of self-organising maps, with units and descriptions.}
    \begin{tabulary}{\textwidth}{llL}
        \hline
        Systematic-tracer variable & Unit & Description \\
        \hline
        \hline
        \ttt{MU\_THRESHOLD (MU)} & mag / arcsec$^{2}$ & $r$-band detection threshold above background; the minimum surface brightness of objects detected after background-subtraction. Fainter objects will be lost from an area of observation where the detection threshold is brighter. \\
        \ttt{psf\_fwhm} & arcsec & Full-width at half-maximum of the $r$-band point-spread function, averaged over $n_{\rm{side}}=512$ ($47\,\rm{arcmin}^{2}$) {\sc{HEALPix}}\tnote{1} \, \citep{Gorski2005} pixels and then interpolated to galaxy locations. The blurring of small, faint sources could cause objects to drop below detection thresholds. \\
        \ttt{psf\_ell} & dimensionless & Ellipticity ($1-q$, where $q$ is the 2D major/minor axis ratio) of the $r$-band point-spread function, also averaged over $n_{\rm{side}}=512$ pixels and interpolated to galaxy locations. A PSF ellipticity indicates non-isotropic blurring of object isophotes, creating challenges for shape estimation and inducing a directional dependence for detections. \\
        \ttt{MAG\_LIM\_x} & mag & \ttt{x}-band limiting magnitude ($5\sigma$ above background in a $2''$ aperture) at the object's location, for each of the 9 bands observed by KiDS-VIKING. KiDS object detection is performed in the $r$-band. \\
        \ttt{EXTINCTION\_r} & mag & Galactic dust extinction in the $r$-band, derived using the \cite{Schlafly2011} coefficients for the \cite{Schlegel1998} dust maps. Dust preferentially scatters short-wavelength light from extragalactic objects; the loss of flux could prevent detections, and the modification of galaxy spectral energy distributions (SEDs) poses other problems, \eg for photo-$z$ estimation. \\
        \ttt{gaia\_nstar} & count / arcmin$^{2}$ & Number density of Gaia DR2 \citep{Arenou2018} $14<G<17$ stars in $n_{\rm{side}}=512$ pixels, interpolated onto galaxy locations. High stellar densities can hamper detections as the light from stars obscures background objects, and can also result in spurious galaxy-detections through the misidentification of PSF-blurred, or blended, stars as galaxies. \\
        \hline
    \label{kids:tab:housekeeping_variables}
    \end{tabulary}
\begin{tablenotes}
\item[1] {\small{\url{http://healpix.sourceforge.net}} \citep{Gorski2005}}
\end{tablenotes}
\tablefoot{Systematic-tracer variables chosen from the KiDS DR4 data products, or from other public data \eg Gaia $n_{\rm{star}}$ . These are variables thought to trace phenomena that may impact upon the observed number density of galaxies. See \cite{Kuijken2019} for details on threshold, PSF, and magnitude-limit parameters. Maps of each of these quantities are displayed in Figs. \ref{kids:fig:Apars} and \ref{kids:fig:BCpars}. Where specified, pixelated systematic-tracer values are interpolated from pixel centres onto galaxy locations via a two-dimensional (RA/DEC) nearest-neighbour interpolation.}
\end{threeparttable}
\end{table*}

Angular clustering correlations are typically measured in bins of galaxy redshift, so as to constrain the galaxy bias of redshift samples, thus combining powerfully with lensing probes such as GGL \citep[\eg][]{Yoon2019}, and to assess the growth of large-scale structure over cosmic time. Accurate redshifts (typically from spectroscopy) are required for the optimal binning of galaxies and modelling of correlations, but these are expensive to obtain for large samples of galaxies over a wide area. Consequently, such wide-field surveys rely upon photometric redshifts (photo-$z$), estimated from broadband photometry (such as the nine filters used in KiDS-$1000$) that sample the spectral energy distributions (SEDs) of the galaxies. In this work, we focus on a subsample of KiDS-$1000$ with high-quality photo-$z$ estimates: the $\sim1$M bright galaxy subsample (with $r\lesssim20$), whose photo-$z$ are computed using {\sc{ANNz2}} neural networks \citep{Sadeh2015}, trained on spectroscopically observed galaxies from the Galaxy And Mass Assembly (GAMA) survey \citep{Driver2009}. This approach to photo-$z$ estimation for bright KiDS galaxies was originally presented by \cite{Bilicki2018} using only optical ($ugri$) photometry from KiDS DR3. In our work we use the updated DR4 bright-sample described in \cite{Bilicki2021}, which leverages the expanded nine-band photometric dataset to achieve photo-$z$s with a typical accuracy of $\sigma_{z_{\rm{phot.}}}\sim0.02(1+z)$ in terms of the normalised median absolute deviation (nMAD).

\begin{figure}
    \centering
        \includegraphics[width=\columnwidth]{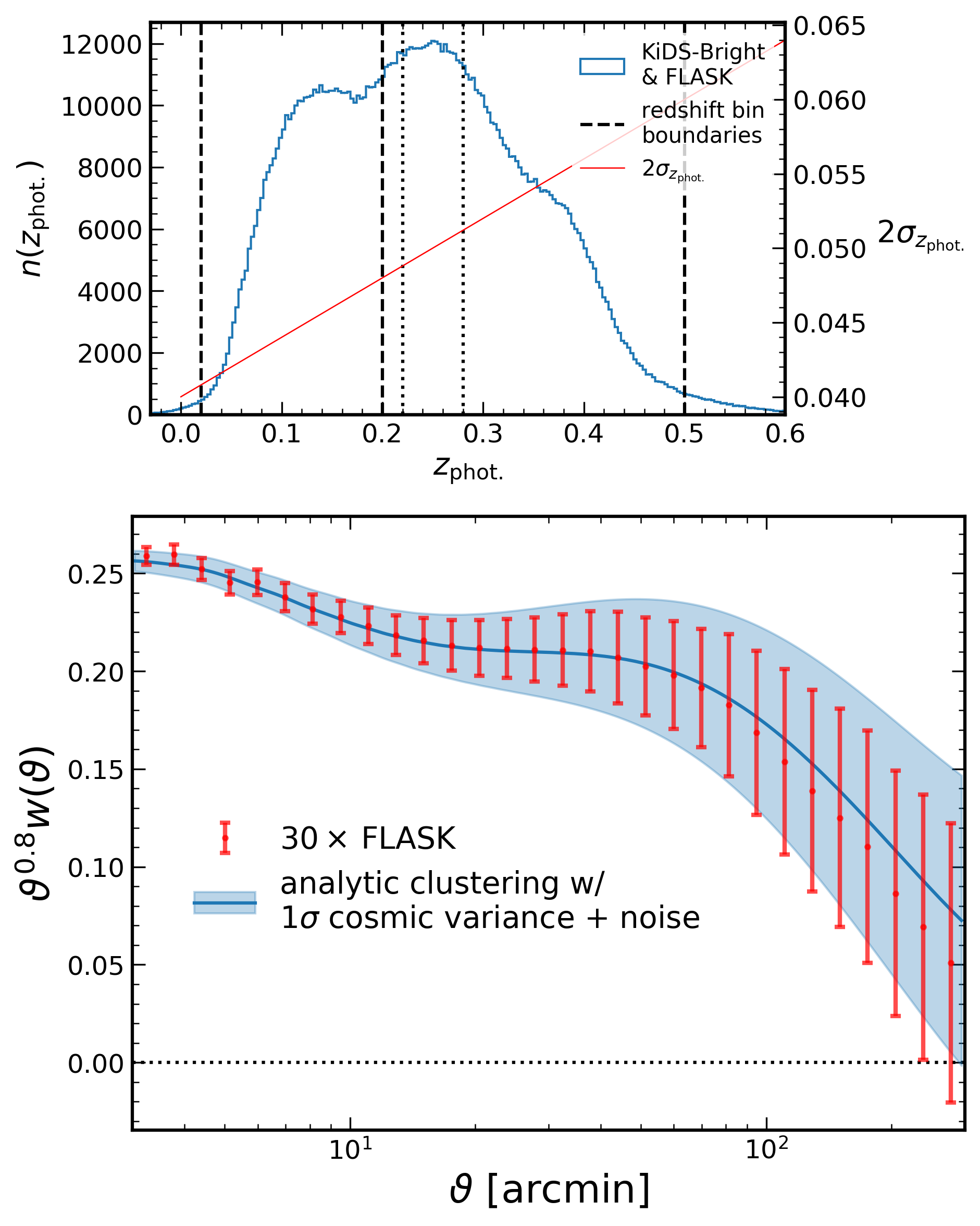}
    \caption[KiDS and {\sc{flask}}: photometric redshift distribution of KiDS bright sample, used to generate \flask catalogues, and consistency of measured \flask clustering with input theory]{\emph{Top: }Photometric redshift distribution of the KiDS-Bright GAMA-like KiDS DR4 bright sample \citep{Bilicki2021}, with the $z_{\rm{phot.}}=\{0.02, 0.2, 0.5\}$ redshift bins (dashed lines) employed in our clustering analysis (Sec. \ref{kids:sec:kids_clustering}) and the additional bins $z_{\rm{phot.,1a}}=\{0.02,0.22\}$ and $z_{\rm{phot.,2a}}=\{0.28,0.5\}$ (dotted lines) defined to have minimal photo-$z$ overlap, as reckoned by the 95\% scatter (red line); $2\sigma_{z_{\rm{phot.}}}\sim0.05$, at that redshift. We use this $n(z_{\rm{phot.}})$ distribution to generate angular power spectra and \flask mock galaxy catalogues. \emph{Bottom: }The full redshift-range angular clustering (red) averaged over \nflask independent \flask lognormal random field realisations of the input power spectrum, which is displayed in blue (see Eq. \ref{kids:eq:wth_theory}) with the theoretical $1\sigma$ error for a KiDS-Bright-like survey, with $0.36$ galaxies per square arcminute over $900\sqdeg$. The galaxy bias is set to unity. Error bars are the root-diagonal of the covariance across the \nflask realisations. We note that for the 30 data points shown, the covariance over just \nflask realisations of the field is quite noisy. We are clearly able to recover the analytical input cosmology with these \nflask realisations from {\sc{flask}}.
    }
    \label{kids:fig:flask_wth}
\end{figure}

The photometric redshift distribution $n(z_{\rm{phot.}})$ for this GAMA-like photometric sample is shown in the top panel of Fig. \ref{kids:fig:flask_wth}. Following \cite{VanUitert2018}, we define two redshift bins for our GAMA-like sample, with edges at $z_{\rm{phot.}}=\{0.02,0.2,$ and $0.5\}$, within (and between) which we measured angular clustering correlations. For the purpose of additional testing, we defined two additional bins within the range of $z_{\rm{phot.}}=\{0.02$ and $0.5\}$, but with inner boundaries well separated in photo-$z$, at $z_{\rm{phot.}}=\{0.22$ and $0.28\}$. This separation minimises overlap between the true redshift distributions in the bins, which are induced by photo-$z$ scatter. We display the approximate 95\% scatter $2\sigma_{z_{\rm{phot.}}}\sim0.04(1+z)$ as a red line, with values on the right-hand axis. We used these GAMA-like KiDS data, `KiDS-Bright' henceforth, for intermediate tests and to assess the ultimate performance of our organised randoms (Sec. \ref{kids:sec:datadriven_systematics} and \ref{kids:sec:kids_clustering}).

\tr{
In our companion letter, \tbf{Wright et al., (in prep.)}, we also explore an application of our organised randoms to measurements of galaxy clustering in the KiDS-$1000$ shear sample. This `gold' sample \citep[see][for details]{Wright2020,Giblin2020,Hildebrandt2020} is $\sim5$ magnitudes deeper than KiDS-Bright, and a factor $\sim20$ more dense on-sky. This increased statistical power should allow for a more faithful sampling of the multivariate systematics-density relation that is hidden in the data, which should also be easier to disentangle from cosmic structure as the faint data are more heavily biased. We expect the performance of organised randoms to \emph{\textup{improve}} on application to faint datasets, posing intriguing possibilities for the future of deep galaxy clustering analyses.
}

\subsection{FLASK}

{\sc{flask}} \citep{Xavier2016} is a public code designed to simulate lognormal (or Gaussian) random fields on the celestial sphere, with configurable tomography and preservation of all relevant correlations between galaxy density and \hlstop{weak lensing} convergence fields, to the sub-percent level. We estimate the error on $w(\vartheta\geq3\,\rm{arcmin})$ to be $\gtrsim2.5\%$ for KiDS-Bright-like statistics from sample variance and Poisson noise considerations alone, hence $<1\%$ accuracy from \flask is sufficient for our purposes.

For the cosmology specified at the end of Sec. \ref{kids:sec:intro} and the $n(z_{\rm{phot.}})$ displayed in the top panel of Fig. \ref{kids:fig:flask_wth}, we computed a `truth' angular power spectrum $C_{\ell}$ with which we used \flask to generate many mock galaxy catalogues from lognormal random fields; these form the basis of initial testing for our organised randoms, as is described in Secs. \ref{kids:sec:artificial_systematics} and \ref{kids:sec:datadriven_systematics}. The bottom panel of Fig. \ref{kids:fig:flask_wth} demonstrates that measurements of \wth in our mocks reliably recover the analytical input clustering and sample variance (plus shot-noise) over \nflask realisations (see Eqs. \ref{kids:eq:wth_landyszalay} and \ref{kids:eq:wth_theory}), modulo noise. These statistics are for \flask realisations with average galaxy densities $0.36\,{\rm{arcmin}}^{-2}$ (\ie the same as KiDS-Bright) simulated within the true KiDS-Bright masked survey footprint, and with a galaxy bias equal to unity. We used these KiDS-Bright-like \flask catalogues, as well as simpler $10\,\rm{deg}\times100\,\rm{deg}$ windows (straddling the celestial equator, to loosely mimic the KiDS survey geometry) initialised with $1$ galaxy $\rm{arcmin}^{-2}$ to test our self-organising maps and random galaxy catalogues.


\section{Self-organising maps}
\label{kids:sec:som}

Self-organising maps\citep{Kohonen1990} are a class of unsupervised neural network methods, designed to project high-dimensional data onto (typically) two-dimensional maps that preserve the topological features of the input space. Proximity on the map therefore tends to denote proximity within the high-dimensional space. SOMs are fast, simple, and useful for problems benefiting from dimensionality reduction, unsupervised classification, and ease of data visualisation. Their use within cosmology has included object selection and classification for large datasets \citep{Geach2012}, template photo-$z$s \citep{Speagle2015b,Speagle2015a}, characterisation of galaxy properties from observables \citep{Davidzon2019}, and calibration of the colour-redshift relation \citep{Masters2017,Masters2019}, enabling direct photo-$z$ calibration \citep{Buchs2019,Wright2020,Hildebrandt2020}.

This work makes use of SOMs for dimensionality reduction and unsupervised classification to identify on-sky areas of observation (or simulation) with correlated observing conditions, as indicated by various systematic survey variables. These variables, such as those describing atmospheric effects and Galactic foreground properties, ought to correlate with phenomena causing systematic alterations to the observed galaxy number density in a wide-field survey. The variables define a systematics-vector\footnote{These are simply rows in a galaxy catalogue, where each element is a number describing the amplitude of some potential source of systematic error at the location of the galaxy.} $\Vec{V}$ per galaxy, which jointly describe an $\mathcal{R}^n$ [where $n=\ttt{length}(\Vec{V})$] dimensional space to be mapped by the SOM. While this space can hypothetically consist of all derived data within the galaxy catalogue, in practice it is beneficial to select variables from the data products available that are likely to trace the true galaxy density variations, so as to allow the maximum density-variation information to be encoded in the SOM.  We therefore explore different choices of tracers that we detail in Table \ref{kids:tab:housekeeping_variables}.

\begin{table*}
    \centering
    \caption[Analysis choices for various self-organising map/randoms-creation configurations: SOM dimensions, number of hierarchical clusters, resolution on-sky, systematic-tracers for training]
    {Analysis choices for various self-organising map and randoms-creation configurations: setup identifier, number of hierarchical clusters, resolution on-sky, and systematic-tracers for training.}
    \label{kids:tab:SOMs}
    \def\arraystretch{1.1}
    \begin{tabular}{lccl}
        \hline
        setup ID &  $N_{\rm{HC}}$ & res (smth) $[\rm{arcmin}]$ & systematics \\
        \hline
        \hline
        \ttt{T1mock} & $100$ & $5\quad(0)$ & \ttt{A1,A2,A3,B1,B2,B3,C1,C2} \\
        
        \ttt{100A} & $100$ & $2.8\quad(0.1)$ & \ttt{MU,psf\_ell,psf\_fwhm} \\
        
        \ttt{800A} & $800$ & $2.8\quad(0.1)$ & \ttt{MU,psf\_ell,psf\_fwhm} \\
        \ttt{800Ares2} & $800$ & $2\quad(0)$ & \ttt{MU,psf\_ell,psf\_fwhm} \\
        
        \ttt{100B} & $100$ & $2.8\quad(0.1)$ & \ttt{MU,psf\_ell,psf\_fwhm,MAG\_LIM\_r} \\
        \ttt{800B} & $800$ & $2.8\quad(0.1)$ & \ttt{MU,psf\_ell,psf\_fwhm,MAG\_LIM\_r} \\
        
        \ttt{100C} & $100$ & $2.8\quad(0.1)$ & \ttt{MU,psf\_ell,psf\_fwhm,MAG\_LIM\_r,gaia\_nstar,EXTINCTION\_r} \\
        
        \ttt{800C} & $800$ & $2.8\quad(0.1)$ & \ttt{MU,psf\_ell,psf\_fwhm,MAG\_LIM\_r,gaia\_nstar,EXTINCTION\_r} \\
        
        \hline
    \end{tabular}
\tablefoot{Various configurations for our SOM and organised randoms-creation pipeline. Throughout the text, we refer to these by their setup ID. This shorthand typically just indicates the number of hierarchical clusters defined on the SOM, \eg \ttt{800} or \ttt{100}, and the number of systematic-tracer variables used in training: \ttt{A} is just the PSF and threshold parameters, \ttt{B} then includes the $r$-band magnitude limit, and \ttt{C} further includes Galactic dust extinction and stellar density (all described in Table \ref{kids:tab:housekeeping_variables}). The exception \ttt{T1mock} refers to the SOM trained against artificial systematic-tracers \ttt{A1-3,B1-3,and C1-2}, which are described in Sec. \ref{kids:sec:artificial_systematics_design}. $N_{\rm{HC}}$ is the number of hierarchical clusters to be defined on a given SOM, `res (smth)' are the resolution and Gaussian smoothing width in arcminutes of the sky-grid used to group galaxies and populate random fields, and `systematics' refers to the set of systematic-tracer variables presented to each SOM in training (see Table \ref{kids:tab:housekeeping_variables}). All SOMs employed in our analysis have dimensions $100\times100$.}
\end{table*}

\begin{figure*}[!htpb]
    \centering
    \includegraphics[width=0.91\linewidth]{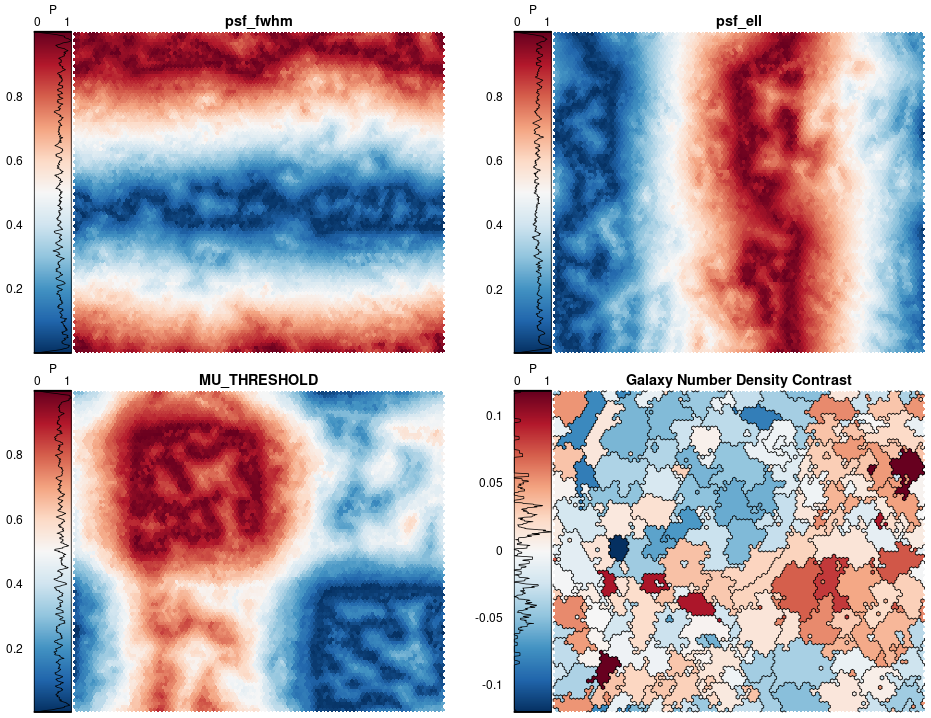}
    \caption[KiDS: self-organised maps of KiDS bright sample systematic-tracers and associated density contrast]{Self-organising maps, with dimension $100\times100$, coloured (\emph{top left, top right, bottom left}) by the systematics values taken on by each cell during the training procedure described in Sec. \ref{kids:sec:som}. In the density contrast panel (\emph{bottom right}), colours and black borders mark the 100 hierarchical clusters (see Sec. \ref{kids:sec:som}) defined according to groupings of cells with similar systematics-vectors for KiDS parameters \ttt{psf\_fwhm,psf\_ell,MU\_THRESHOLD} (Table \ref{kids:tab:housekeeping_variables}). Systematic-tracer variables are linearly mapped onto the interval $[0, 1]$ before being passed to the SOM, hence the colour-bar ranges. The density contrast panel (\emph{bottom right}) maps clusters of SOM pixels from their vectors of systematic-tracer values back onto a relative number density on-sky and reveals almost all systematic density fluctuations to be at $\lesssim10\%$, as reckoned by this SOM configuration (\ttt{100A}; Table \ref{kids:tab:SOMs}).}
    \label{kids:fig:SOMfig}
\end{figure*}

The SOM algorithm starts by instantiating a grid with user-specified dimensions, for example $100\times100$ for a two-dimensional SOM containing $10^4$ cells. Each cell is then assigned a randomised weights-vector $\Vec{W}$ of the same length as the galaxy systematics-vectors, that is, the number of systematic-tracer variables $n$. To train the SOM, galaxy systematics-vectors $\Vec{V}$ are then chosen at random and presented to the SOM lattice. At each step of the training, the SOM cell with weights $\Vec{W}$ most closely matching the training galaxy systematics-vector $\Vec{V}$ is termed the best-matching unit (BMU). The match is typically quantified through the Euclidean distance $d$ between the SOM cell weights-vector $\Vec{W}$ and the galaxy systematics-vector $\Vec{V}$ as

\begin{equation}
    d = \sqrt{\sum^{n}_{i=1} \left(V_{i}-W_{i}\right)^{2}} \quad ,
\end{equation}
where the minimum $d$ over the grid belongs to the BMU. Next we identify SOM cells within some radius $\sigma(t)$ (the neighbourhood) of the BMU, and modify their weights-vectors $\Vec{W}$ to be closer to $\Vec{V}$, with more significant modifications for cells nearer the BMU. The resulting weights-vectors are given by

\begin{equation}
    \Vec{W}(t+1) = \Vec{W}(t) + L(t)\,\Theta(t,\sigma)\left[\Vec{V}(t)-\Vec{W}(t)\right]
,\end{equation}
where $t$ denotes a time-step (\ie the presentation of a new training galaxy to the SOM), the learning rate $L$ sets the strength of modifications, and $\Theta$ implements the distance-to-BMU dependence thereof. The final feature is that all of (i) the radius $\sigma$ within which cell weights-vectors are to be modified, (ii) the learning rate $L$, and (iii) the BMU-distance dependence $\Theta$, are exponentially decaying with each time-step, hence their dependence upon $t$. In this way, the SOM converges to a final representation as the training data are exhausted. Once all galaxies have been presented to the SOM, each cell on the SOM grid carries a weights-vector describing some unique position in the $n$-dimensional systematics parameter space, and the full collection of $10^{4}$ cells spans the entire space sampled by the galaxies.

In our implementation, the resulting 2D map then represents the landscape of possible systematics-vectors realised by the data in question. By computing the distances between points in the space, we can then divide the landscape into $N_{\rm{HC}}$ maximally separated hierarchical clusters \citep[see][for a description of this process]{Wright2020}. Briefly, hierarchical clusters are defined by assigning each SOM cell to its own cluster, and then iteratively combining the two least-separated clusters (in our case by Euclidean distance between the cell weight-vectors $\Vec{W}$) into one, until only a single cluster remains. At each iteration, the cluster centres are recomputed using the Lance-Williams dissimilarity formula \citep[see][]{Defays1977,Iezzi2014}, invoking complete-linkage clustering to generate the most similar clusters.  This iterative process constructs a cell-merger dendrogram that can then trivially be sliced at the desired number of clusters $N_{\rm{HC}}$. Each cluster of cells then contains a unique subset of the total galaxy sample, which is described by similar systematics-vectors $\Vec{V}$. In this way, the combination of the SOM and hierarchical clustering is able to construct $N_{\rm HC}$ distinct groups of sources with similar systematics properties, but which occupy non-contiguous areas of the sky. The average total area on-sky that is spanned by each of the $N_{\rm HC}$ groups is therefore determined by the area of the dataset in question (\ie $\sim900\sqdeg$ for KiDS-Bright) and $N_{\rm HC}$. 

From the on-sky distribution and galaxy-count corresponding to each hierarchical cluster, we can compute an estimate of the galaxy number density associated with each region of the high-dimensional \emph{\textup{systematics space}} (\ie the space of systematic-tracer variables from Table \ref{kids:tab:housekeeping_variables} and Figs. \ref{kids:fig:Apars} and \ref{kids:fig:BCpars}). Our randoms creation algorithm uses this information to populate the survey volume with variably dense but locally unclustered random points. \tbf{Moreover, each random point is constructed as a copy of one of the training galaxies from the hierarchical cluster to which it belongs, carrying all of the parent galaxy physical and photometric properties: a clone. Clones are therefore scattered only within the (again, non-contiguous) areas of sky represented by the parent hierarchical cluster. As a result, galaxy sample selection effects and associated selection of specific systematic clustering patterns are easily reproduced in the randoms catalogue by a simple selection of clones satisfying the chosen galaxy selection function.} Acting as the reference points in a galaxy clustering measurement, the randoms ought to then compensate for density variations correlating with any combination of the tracked systematic variables\hlstop{,} for any selection of galaxies used to train the SOM.

The number of clusters $N_{\rm{HC}}$ is an important tuning parameter for this analysis: in addition to determining the area-per-cluster, $N_{\rm{HC}}$ must also trade off against the amount of discretisation of the systematics space to be mapped. If we invoke too few clusters in our high-dimensional space, each will span a wide range of observational (\ie systematic) and cosmological regimes, and could thus lose discerning power as a result (indeed, $N_{\rm HC}=1$  returns completely uniform randoms by definition). Conversely, invoking too many clusters in the same space will result in too much freedom, and begin to over-fit the number density versus systematics relation. This over-fitting can itself be pathological, as we show below, if the individual clusters begin to trace the cosmic (\ie not systematic) structure. 

Our various choices for SOMs are detailed in Table \ref{kids:tab:SOMs}, and Fig. \ref{kids:fig:SOMfig} displays SOMs of the configuration \ttt{100A} after training on KiDS-Bright. The bottom left, top left and top right panels are coloured by the \ttt{MU,psf\_fwhm,and psf\_ell} (Table \ref{kids:tab:housekeeping_variables}) values for these cells, and the bottom right panel connects them to the galaxy density contrast, $(n_{\rm{gal}}-\langle{n_{\rm{gal}}}\rangle)/\langle{n_{\rm{gal}}}\rangle$, in 100 hierarchical clusters (denoted by discrete patches of colour with black borders). Inspecting any region of the density contrast map, we can easily identify correlations between density contrast and various systematic parameters. For example, we can note that the upper left quadrant of the density contrast SOM is populated by clusters of generally below-average (\ie blue) density contrast. This is clearly correlated with above-average values of the $r$-band imaging surface brightness limit \ttt{MU\_THRESHOLD}. The opposite effect (\ie above-average density contrast) is seen for below-average values of \ttt{MU\_THRESHOLD}, except for clusters where the value of the PSF full width at half maximum (FWHM) is above average; in this case, the density contrast is once again below average. These conclusions are easily drawn from the SOM and are a significant advantage of our methods presented here.

\section{Validating SOMs with artificial systematics}
\label{kids:sec:artificial_systematics}


First, we elucidate the capabilities and limitations of SOMs in recognising galaxy density-systematic correlations through the use of artificial systematic fields with complex correlations to the depletion of the galaxy number density. 

\subsection{Creating artificial systematic density fluctuations}
\label{kids:sec:artificial_systematics_design}

\begin{figure*}[!ht]
    \centering
    \includegraphics[width=\textwidth]{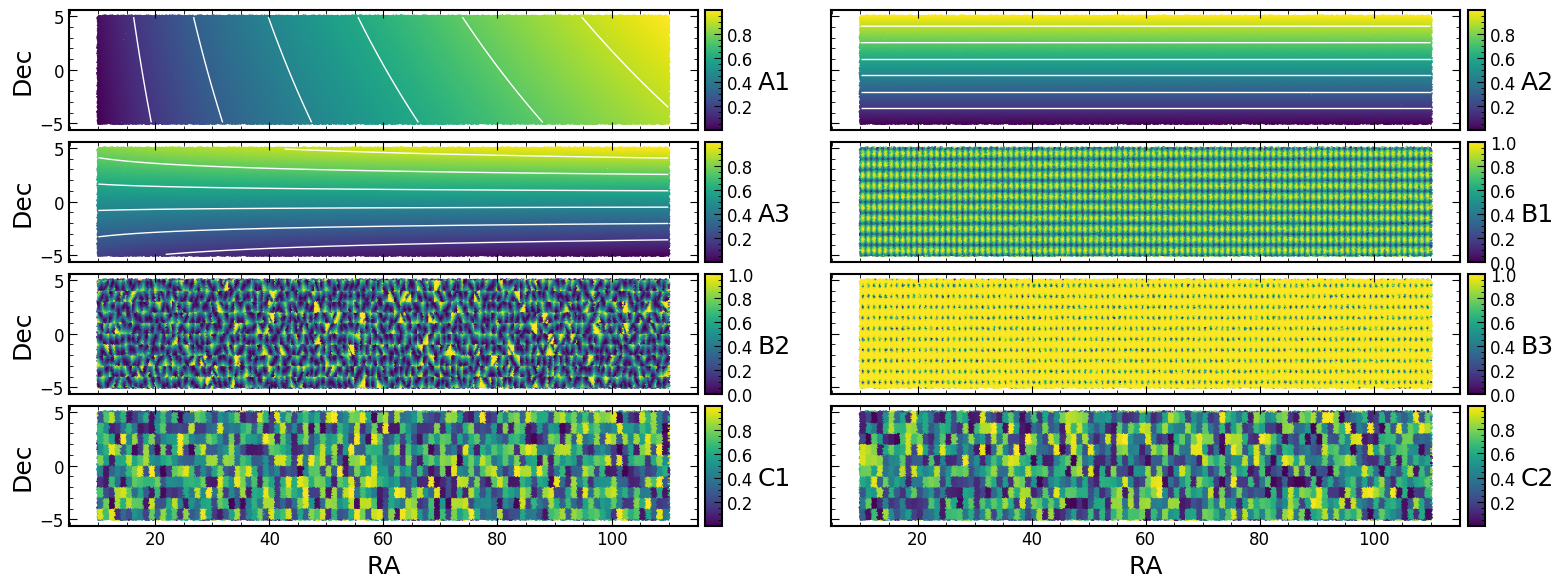}
    \caption[{\sc{flask}}: maps of artificial systematics-variables]{ {\sc{flask}}-generated uniform random field, binned hexagonally (cell scale $\sim10\,\rm{arcmin}$) in RA/DEC and coloured according to our various artificial systematic-tracer variables. \ttt{A}-types vary smoothly over large angles, \ttt{B}-types vary over $1\sqdeg$ tiles with a 2D Gaussian form, and \ttt{C}-types are single-valued for each tile. Iso-contours (white lines) are lain over \ttt{A}-type variables for clarity, and to illustrate the subtle difference between \ttt{A2} and \ttt{A3}.}
    \label{kids:fig:imagpars}
\end{figure*}

Fig. \ref{kids:fig:imagpars} shows on-sky distributions of $1\,\rm{arcmin}^{-2}$ \flask galaxies, colour-coded by our set of eight artificial systematics variables, which were designed to mimic realistic spatial patterns in KiDS-like wide-field observations. \ttt{A}-type variables vary smoothly over large angles, in a manner similar to Galactic foregrounds. \ttt{B}-type variables have a two-dimensional Gaussian form, which varies independently and discretely in $1\times1\sqdeg$ tiles, thereby mimicking telescope and camera effects such as PSF variations over the focal plane. Finally, \ttt{C}-type variables vary discretely between tiles but are constant within them, thereby mimicking per-exposure effects such as limiting depth variations that arise from the use of a step-and-stare observing strategy (meaning that each tile is only observed once, per-band, over the course of the entire survey, thereby increasing sensitivity to the variable sky brightness on any given night). Each effect varies in the range $[0,1]$ for simplicity. By construction, these analytic systematics have no serious outliers.

\begin{figure}
    \centering
    \includegraphics[width=\columnwidth]{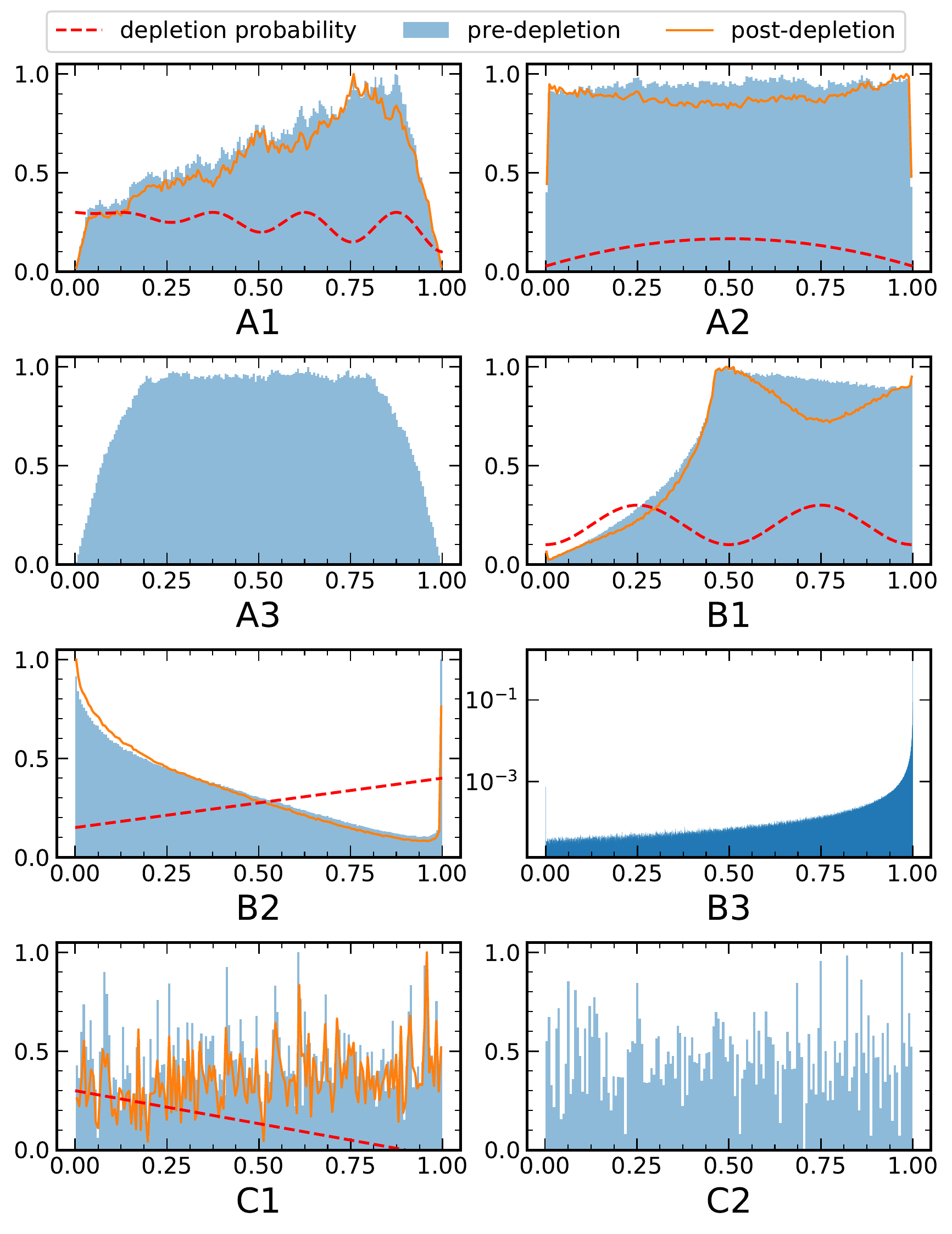}
    \caption[{\sc{flask}}: the probability of depletion as a function of each artificial systematic, with distributions pre- and post-depletion]{Histograms, each normalised to a maximum of unity, of the artificial model systematics (see Fig. \ref{kids:fig:imagpars}) we imposed upon our \flask lognormal random field galaxies, both before (blue) and after (orange) applying our probabilistic depletion functions, shown as red dashed lines. Dummy variables (\ttt{A3,B3,and C2}) have no depletion applied. They are intended to distract the SOMs.}
    \label{kids:fig:depletion_functions}
\end{figure}

To create spurious density modes as multivariate functions of our artificial systematics, we invented an independent \emph{\textup{depletion function}} for each parameter, shown in Fig. \ref{kids:fig:depletion_functions} as red dashed lines. These modes each describe the probability that an object will be discarded from the catalogue as a function of the position \hlstop{of the object} in systematics space, thereby depleting the galaxy number density at that point in the space. We note that some systematic variables cause no depletion (\ttt{A3,B3,and C2}) and act instead as dummy parameters for the SOM to navigate. In Appendix Fig. \ref{kids:fig:excess_deplprob} we show the spatially variable excess probability of depletion resulting from the depletion functions of Fig. \ref{kids:fig:depletion_functions} as applied to the systematics-maps of Fig. \ref{kids:fig:imagpars}.

We applied each depletion function individually\footnote{The case of multiple potentially correlated systematic biases working in confluence is more realistic, but a SOM could then only infer the combined multivariate depletion function, making an assessment of success more complicated. We explore the consequences of multiple biases when we consider data-driven systematic density fluctuations in the next section.} to a set of \nflask \flask mocks. We then trained a SOM using each depleted mock galaxy sample with the \ttt{T1mock} configuration from Table \ref{kids:tab:SOMs} and created a hierarchy of cell clusters on each trained SOM. The number densities of the clusters should then reflect the input depletion functions per individual systematic from Fig. \ref{kids:fig:depletion_functions}. This test is thus a simple verification of the \hlstop{ability of the} SOM to characterise galaxy-density variations as a function of some systematic effect, with a particular pattern on-sky. Furthermore, we can use this test to explore possible systematic patterns that may cause the SOM to fail in its corrective pursuit.

\begin{figure*}
    \centering
    \includegraphics[width=\linewidth]{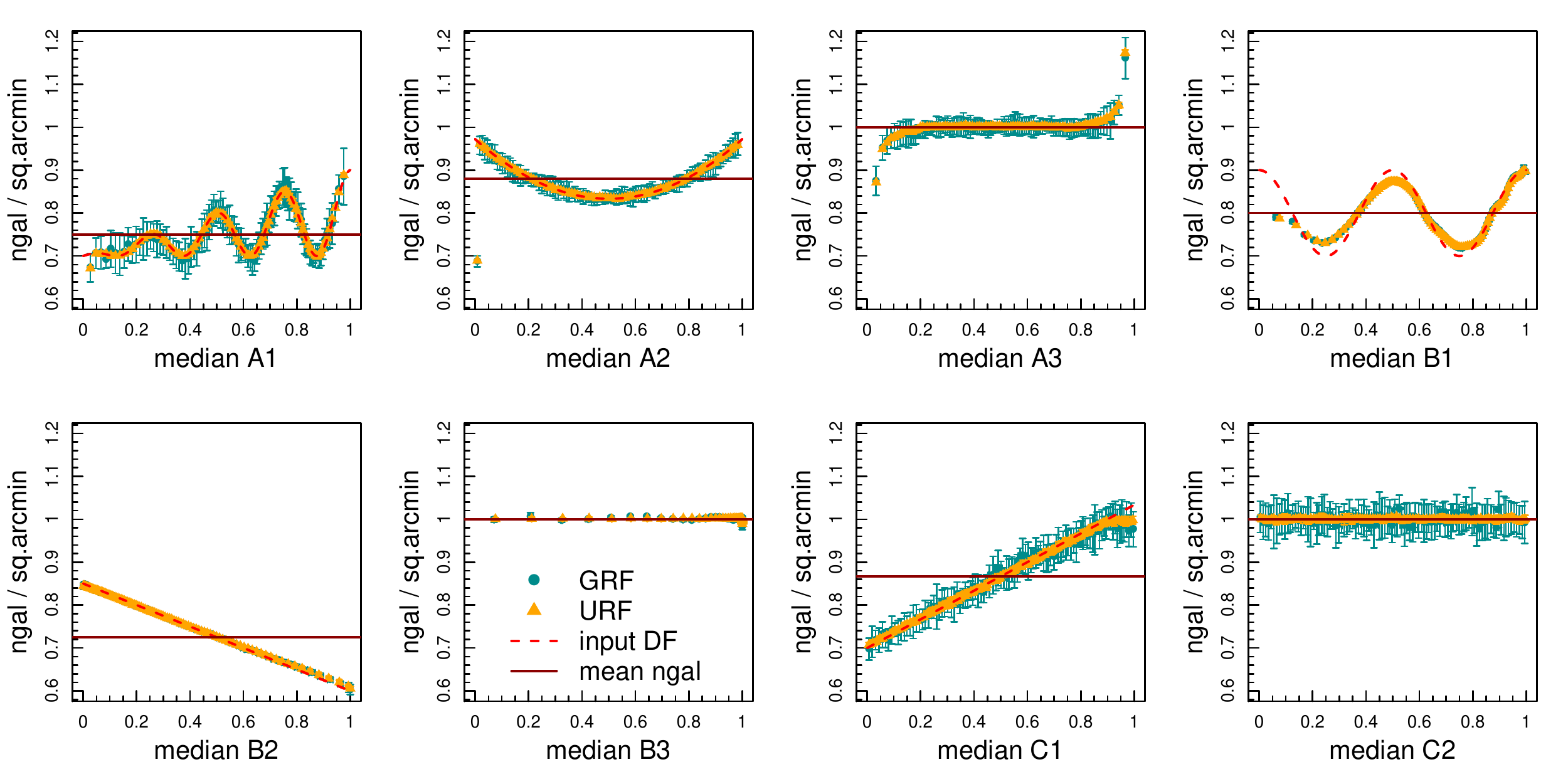}
    \caption[{\sc{flask}}: the self-organising map's recovery of input depletion functions in depleted lognormal and uniform  random fields]{Recovery by the \ttt{T1mock} SOM (Table \ref{kids:tab:SOMs}) of the artificial biases applied to \nflask {\sc{flask}} realisations (see Figs. \ref{kids:fig:imagpars} and \ref{kids:fig:depletion_functions}) of random fields. Panels show the galaxy number density ($\rm{arcmin}^{-2}$) of 100 hierarchical clusters, defined on the SOM, against the median systematic value in each of these clusters. Green points give the results for SOMs trained against depleted lognormal random fields (GRF), and orange points give the same for uniform random fields (URF), each generated using {\sc{flask}}. The functions describing systematic galaxy depletion relative to our artificial systematic parameters (from Fig. \ref{kids:fig:depletion_functions}) are converted into expected number densities and shown here as dashed red lines (`input DF'; depletion function). The solid horizontal lines indicate the global average number densities expected per systematic after depletion. Points where both our URF and GRF data sit systematically away from the expected values are artefacts due to footprint edge-effects in our \flask simulations. They do not affect the conclusions of this test (see Sec. \ref{kids:sec:artificial_syst_results}).}
    \label{kids:fig:test1_tob}
\end{figure*}

\subsection{Characterisation of artificial systematic fluctuations}
\label{kids:sec:artificial_syst_results}

We find that the SOMs are able to recover individual depletion functions with good accuracy. Fig. \ref{kids:fig:test1_tob} displays number densities versus systematic values for hierarchical clusters defined on the \ttt{T1mock} SOM (Table \ref{kids:tab:SOMs}), along with our input depletion functions in red. Error bars are the root-diagonal of the covariance over \nflask realisations, and we display the result for both lognormal (green) and uniform (orange) random fields (GRF and URF respectively). The additional noise seen in $n_{\rm{gal}}$ for the GRF case is due to the simulated cosmic structure, and can be seen to correlate with systematics that localise finite regions on-sky (\ie \ttt{A} and \ttt{C} types). In contrast, \ttt{B}-type parameters show little difference between uniform and lognormal random fields, as individual systematics clusters cover a much less localised on-sky area. They are consequently much less sensitive to (local) variations in cosmic number density. We also verified that the SOM does not create depletion functions where none exist. Appendix Fig. \ref{kids:fig:test1_tou} shows that the SOM recovers the mean galaxy density in all cases when training the SOM on the same parameters but without any depletion of galaxy number densities.

We note and address the presence of some irregularities in the distributions of Fig. \ref{kids:fig:test1_tob}, bearing in mind that the test here is merely to recover the form of depletions, and that these are superseded by two-point statistics in the next section. In extrema, \ttt{A}-type parameters appear to have outlier clusters with very high or low number density. These are artefacts from binning galaxies on the $5\,\rm{arcmin}$ Cartesian grid. At the footprint edges, some grid cells are only partially filled, resulting in seemingly under- or overdense clusters. The \ttt{B1} clusters are also affected by footprint edges, where \ttt{B1} is predominantly low (Fig. \ref{kids:fig:imagpars}), and the amplitude of recovered fluctuations is suppressed with respect to the expectation for decreasing values of \ttt{B1}. The distribution of \ttt{B1} (Fig. \ref{kids:fig:depletion_functions}; blue histogram) drops sharply below values of $\sim0.5$, whilst the probability of depletion rises (Fig. \ref{kids:fig:depletion_functions}; dashed red line), resulting in a thin, sparsely populated, grid-like distribution of objects across the footprint that begins to noisily sample the mean density and causes further suppression of inferred fluctuations when the periodicities of the \ttt{B1} fluctuation and the on-sky grid are misaligned. We verified that such effects can be mitigated by modifying the on-sky grid used to discretise the galaxy sample, but we caution that a reduction in the grid size can cause artificial masking of empty patches of sky, given sparse galaxy data. In practice, these effects can be mitigated by ensuring mutual masking of the randoms and galaxy catalogues, and by tuning the on-sky grid size and grid-smoothing parameters to account for the range of number densities in the dataset under investigation. \tbf{We further note that these artefacts are also seen in Appendix Fig. \ref{kids:fig:test1_tou}, demonstrating that they are not related to the applied systematic density fluctuations.}

Given our experience with this test, we settled on a fiducial setup with an on-sky resolution of $2.8\,\rm{arcmin}$, and invoked a Gaussian smoothing kernel with a standard deviation of $0.1\,\rm{arcmin}$. This allows empty grid-cells on the sky to be affected by their populated neighbours, thus ensuring that the randoms respect the survey mask even for sparsely distributed galaxy data (such as the KiDS-Bright sample).

\section{Validating organised randoms with data-driven systematics}
\label{kids:sec:datadriven_systematics}

\tbf{The ultimate goal for our organised randoms is to provide a simple and unsupervised method with which to debias galaxy clustering signals upon measurement. In the previous section, we demonstrated that self-organising maps are capable of identifying systematic density fluctuations in data. We now use the SOMs to infer realistic systematic fluctuations from KiDS-Bright data and apply them to simulated \flask catalogues. Training organised randoms against these biased data, we can then assess their ability to recover unbiased two-point statistics.}

We first note that many typical approaches to systematics-handling for galaxy clustering feature the correction of pixel density-systematics one-point correlations \citep[\eg][]{Suchyta2016,Elvin-Poole2017,Rezaie2019,Kitanidis2019,Vakili2020}. \tbi{We find these metrics to be unsuitable, at least in isolation, for validation of our own methods, where a strict correction of individual pixel density-systematic correlations is not necessary for the recovery of unbiased clustering two-point functions, and may be a misleading measure of performance. A perfect performance with regard to one-point density-systematic correlations could mask a total failure of our methods at the two-point clustering level}, featuring a dramatic suppression of real clustering signals. For further details, see Appendix \ref{kids:app:1point_correlations}.

\subsection{Galaxy clustering}
\label{kids:sec:clustering}

For all measurements of galaxy clustering two-point correlations, we considered the angular correlation function $w(\vartheta)$, and made use of the public software {\sc{TreeCorr}}\footnote{\url{https://github.com/rmjarvis/TreeCorr}} \citep{Jarvis2004}. The standard Landy-Szalay \citep{Landy1993} estimator for angular galaxy clustering is given as

\begin{equation}
    \hat{w}^{ij}(\vartheta) = \frac{\Theta(\vartheta) \left\{D_{i}D_{j} - D_{i}R_{j} - R_{i}D_{j} + R_{i}R_{j} \right\}}{\Theta(\vartheta) \left\{R_{i}R_{j}\right\}} \, \quad ,
    \label{kids:eq:wth_landyszalay}
\end{equation}
where $i,j$ denote the galaxy samples being correlated\hlstop{;} photo-$z$ bins, in this work. $DD$, $DR,$ and $RR$ denote normalised\footnote{Normalisation is by the product of total counts $N_{i}N_{j}$ for samples $i$ and $j$, that is, the count of all possible pairs. For auto-correlations, $N$ is subtracted from the product, as a galaxy cannot be paired with itself.} pair-counts between galaxies $D$ and randoms $R$, and the bin-filter operator $\Theta(\vartheta)$ rejects pairs with separations falling outside of the angular bin centred at $\vartheta$. We considered both auto- ($i=j$) and cross-correlations ($i\neq{}j$) when evaluating our organised randoms.

Assuming a flat universe and a linear galaxy bias, and under the Limber and flat-sky approximations, we obtain a theoretical expectation for $w(\vartheta)$ between galaxy samples $i$ and $j$ as follows \citep{Limber1953,Loverde2008}:

\begin{equation}
 \label{kids:eq:wth_theory}
 \begin{split}
    w^{ij}(\vartheta) = &\,\, 2\pi\, b^{i}_{\rm{g}}b^{j}_{\rm{g}} \,\, \times \\ & \int^{\infty}_{0} \frac{\rm{d}\ell}{\ell} J_{0}(\ell\vartheta)  \int^{\chi_{\rm{h}}}_{0} {\rm{d}}\chi \, \frac{p^{\, i}(\chi)p^{\, j}(\chi)}{\chi^{2}} \, P_{\delta}\left(\frac{\ell+1/2}{\chi}, \chi\right) \quad ,
 \end{split}
\end{equation}
where $b_{\rm{g}}$ denotes the linear galaxy bias, $\ell$ the angular frequency, $J_{0}$ the zeroth-order Bessel function of the first kind, $\chi$ the comoving distance to a given redshift, $p(\chi)$ the normalised comoving distance distribution of the sample, and $P_{\delta}$ the matter power spectrum with non-linear corrections \citep[for which we employed the \ttt{halofit} model of][]{Smith2003,Takahashi2012}. The right-most integral is equal to the (Limber-approximate) angular power spectrum $C_{\ell}$.

\tbi{We note here that an alternative approach to the construction of organised randoms, which correct the clustering through the $R$ terms in the above estimator (Eq. \ref{kids:eq:wth_landyszalay}), would be to correct the galaxy densities directly, for example using galaxy weights, and then employ standard, uniform randoms in the estimator \citep[as in][]{Ross2011,Elvin-Poole2017,Rezaie2019,Kitanidis2019,Vakili2020,Wagoner2020}. Our SOM methods lend themselves naturally to this approach, and we intend to explore this with angular power spectra in an upcoming KiDS analysis.}

\subsection{Realistic biasing of \flask fields}
\label{kids:sec:flask_biasing}

\begin{figure*}
    \centering
    \includegraphics[width=\linewidth]{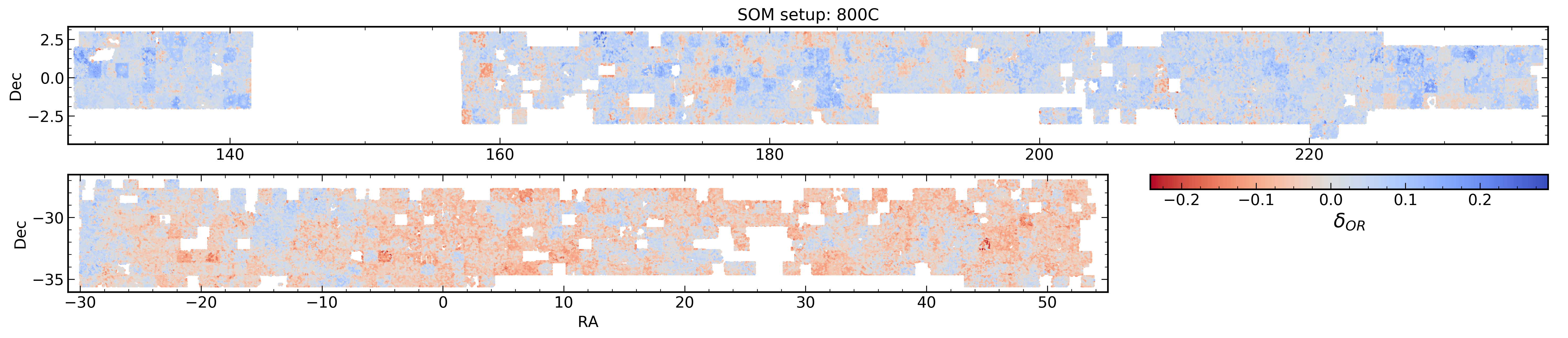}
    \caption[KiDS: systematic density modes across the survey footprint, as inferred by the self-organising map]{Systematic density contrast $\delta_{\rm{OR}}$ (Eq. \ref{kids:eq:Pdepletion}) inferred for the KiDS-North (\emph{top panel}) and KiDS-South (\emph{bottom panel}) areas of the KiDS-$1000$ bright sample (KiDS-Bright) by the \ttt{800C} SOM setup. We use these maps to construct our organised randoms, populating the footprint to mirror the systematic density modes.}
    \label{kids:fig:deltaOR}
\end{figure*}

An application of SOMs to real data lacks a clear notion of truth: we do not know exactly how our tracer variables relate to the deprecation of observed galaxy densities. Thus we cannot know whether the systematic density relations inferred by the SOMs are contaminated by cosmological density variations; for example, it could be that a real, local north-to-south galaxy density gradient happens to correlate with some systematic-tracer. The SOMs could \hlstop{then} falsely identify such a variable to be tracing the \emph{\textup{source}} of the density gradient. Any randoms organised accordingly would then act to erase real cosmological clustering signals. Moreover, we do not know the true underlying clustering signal: if our corrections were already sufficient to recover the unbiased clustering, but one-point pixel density-systematics correlations \citep[\eg][see Appendix \ref{kids:app:1point_correlations}]{Elvin-Poole2017,Rezaie2019} still revealed correlations with systematics, how would we know to stop?

Here we assess the performance of organised randoms in recovering unbiased clustering signals, and also address the question of over-fitting to spatial, possibly cosmological trends in galaxy density. For the reasons detailed above, this is difficult to do for real galaxy data such as KiDS-Bright, where any biases are already present. We sidestepped these concerns with synthetic galaxy distributions from {\sc{flask}}. The test method is as follows: we fed real KiDS-Bright galaxies and systematics data to a SOM and inferred the spatial pattern of depletion. It is unimportant whether the pattern is contaminated by LSS at this stage. We performed a nearest-neighbour interpolation in RA/DEC to port the real spatial distributions of systematics from KiDS-Bright onto \flask galaxies simulated within the same mask. We then applied the SOM-inferred systematic density fluctuations to the mocks, probabilistically, as

\begin{equation}
    P_{\rm{depl.}}(\vec{x}) = 1 - \frac{n_{\rm{gal}}}{n'_{\rm{gal}}} \big[1 + m\,\delta_{\rm{OR}}(\vec{x})\big] \quad ,
    \label{kids:eq:Pdepletion}
\end{equation}
where $P_{\rm{depl.}}(\vec{x})$ is the probability that a galaxy at position $\vec{x}$ will be lost; a uniform random draw in the range $[0,1]$ must exceed $P_{\rm{depl.}}(\vec{x})$ for the galaxy to be retained. $n_{\rm{gal}}$ and $n'_{\rm{gal}}$ are the target and initial \flask number density. We can only remove galaxies from the mocks, so we initialised the \flask realisations with $n'_{\rm{gal}}=0.72$ galaxies $\rm{arcmin}^{-2}$ and then generated systematic under- or overdensities with respect to the mean (target) KiDS-Bright density of $n_{\rm{gal}}=0.36\,\rm{arcmin}^{-2}$. $\delta_{\rm{OR}}(\vec{x})$ is the density contrast at position $\vec{x}$ sourced by systematics\footnote{This inferred density contrast from systematics forms the basis for our organised randoms generation algorithm, although for this test we created organised randoms only after training separate SOMs against the depleted \flask mocks.} according to the SOM, and $m$ is a scalar variable that we can use to manually modify the amplitude of the applied depletion whilst retaining its functional relation to the KiDS-Bright systematics distribution. Taking $m>1$ would intensify the depletion relative to that present in KiDS-Bright, whilst $m<1$ would yield a comparatively soft depletion. We display an example map of $\delta_{\rm{OR}}$, inferred from KiDS-Bright galaxies, in Fig. \ref{kids:fig:deltaOR}.

Repeating this procedure for many realisations of the underlying cosmology, we created the truth case of a single, global pattern of galaxy depletion. Running the SOMs again, now against the depleted \flask mocks, we assessed how consistently they are able to recover this truth for many different realisations of the constant cosmological background. If the SOMs are able to retrieve the fixed systematic depletion pattern from many independent realisations of the cosmic structure, then we can assert that the inferred $\delta_{\rm{OR}}(\vec{x})$ is uncontaminated by cosmology.

Having thus created many realisations of KiDS-Bright-like data for which we can turn systematic biases on or off, we ran our SOMs and assess the corrective performance of our organised randoms with various measurements of $w(\vartheta)$. A caveat is that our depletion of the \flask galaxies was derived from runs of the SOM against KiDS-Bright, thus the bias can be said not to be \emph{\textup{entirely}} realistic (although far more so than the artificial systematics of the previous section, for which the accuracy of capture was excellent) but dependent upon the configuration of the initial SOM. We therefore considered many different configurations along with modifications (through the $m$ parameter from Eq. \ref{kids:eq:Pdepletion}) to the intensity of depletion that we applied to the \flask mocks.

Over the course of testing, we recognised that systematic-tracer set \ttt{B} (Table \ref{kids:tab:housekeeping_variables}) was a relatively uninformative yardstick between sets \ttt{A} and \ttt{C}; therefore we limit our discussion from here to parameter sets \ttt{A and C}, which are instructive for our work with the relatively unbiased KiDS-$1000$ bright sample. \tbi{We note also that the soft systematic biases present in KiDS-Bright may well be treatable with a simpler linear regression formalism \citep[\eg][]{Ross2011,Elvin-Poole2017}, although we have not tested this claim. Our more complex SOM-derived model is, however, more likely to be necessary for our work with the faint KiDS shear sample (Wright et al., in prep.), which features stronger systematic biases.}

\subsection{Correction of data-driven systematic fluctuations}
\label{kids:sec:test3_results}

We devised a battery of \wth tests to assess the performance of our organised randoms at the two-point level. The key variables that we changed were (i) the parameters used to train a SOM against KiDS-Bright and (ii) the number of hierarchical clusters $N_{\rm{HC}}$ defined on that KiDS-Bright SOM. These determine the spatial depletion to be ported onto {\sc{flask}} realisations, as described in Sec. \ref{kids:sec:datadriven_systematics}. We also changed (iii) the parameters used to train a second SOM against biased \flask realisations and (iv) the number of hierarchical clusters $N_{\rm{HC}}$ defined on that \flask SOM. This allowed us to break any circularity by comparing independent SOMs. Finally, we also experimented with (v) intensifying the depletion through the $m$ parameter (Eq. \ref{kids:eq:Pdepletion}).

\begin{figure*}[!htp]
    \centering
        \includegraphics[width=\linewidth]{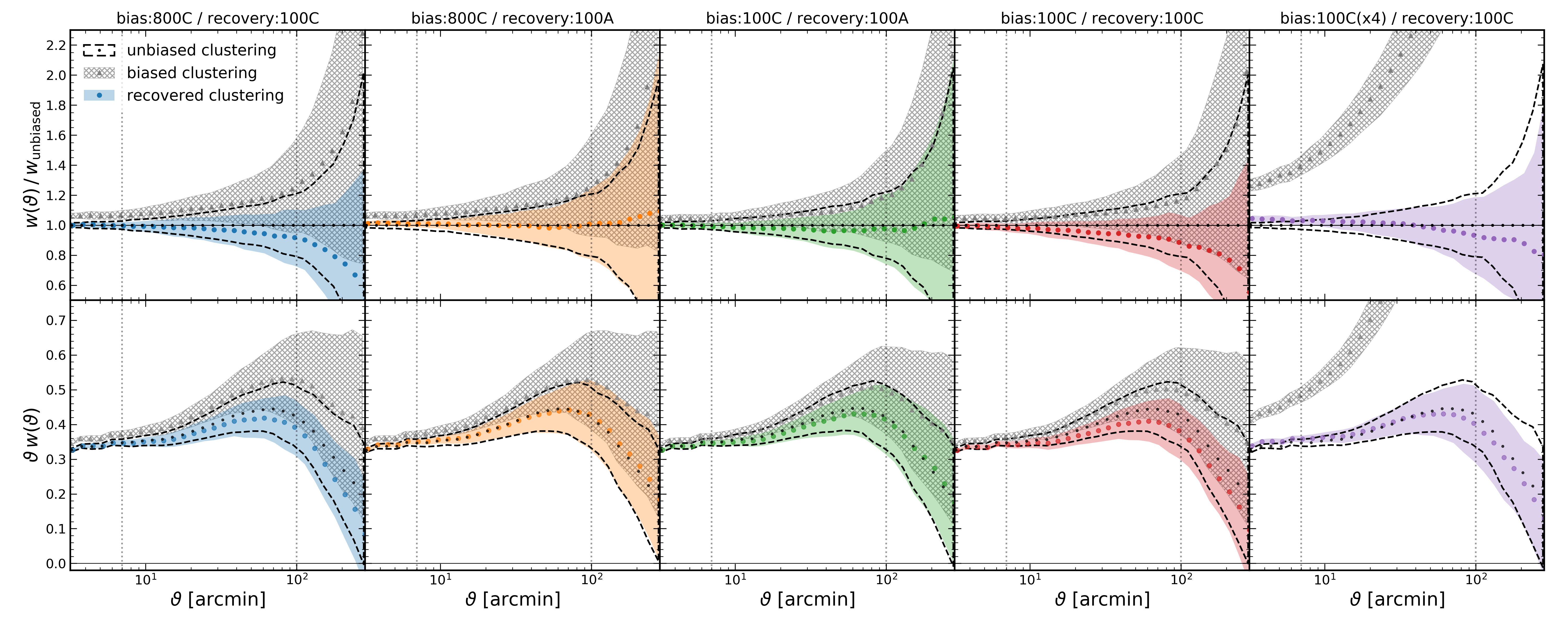}
    \caption[{\sc{flask}}: \wth signals in biased \flask fields, with/out corrective randoms -- best performance]{Angular clustering correlation functions \wth measured in \flask fields after they have been depleted (as described in Sec. \ref{kids:sec:flask_biasing}; Eq. \ref{kids:eq:Pdepletion}) according to the output of various SOMs trained against the KiDS-$1000$ bright sample (KiDS-Bright). For each configuration, panel titles give the bias:SOM trained against KiDS-Bright, and the recovery:SOM trained against biased \flask data to create organised randoms. \emph{Top:} Ratios of measured clustering signals to the true, unbiased clustering. \emph{Bottom:} Unbiased angular clustering signature (black points and dashed curves; measured with uniform randoms on unbiased \flask fields), compared with biased (grey triangles and hatching; measured with uniform randoms) and recovered (coloured points and shading; measured with organised randoms) clustering signals measured in the depleted fields.
 Errors are the root-diagonal of the covariance over \nflask \flask realisations, and all are given to $\pm{1}\sigma$. From left to right, the \flask fields are biased according to $\delta_{\rm{OR}}$ from the \ttt{800C,800C,100C,100C,and 100C} SOMs, trained against KiDS-Bright. The $m$-parameter (Eq. \ref{kids:eq:Pdepletion}) is set to unity in all but the purple panel (where $m=4$). \hlstop{Each correlation measured in the depleted fields using organised randoms displays improved consistency with the unbiased signal, compared to the signals measured using uniform randoms.}
    }
    \label{kids:fig:t3_master}
\end{figure*}

\tbf{
We find that the performance and flexibility of our organised randoms are excellent. The randoms were able to consistently mitigate biases in \flask fields, even when they had limited sensitivity to smaller scales (through reduced $N_{\rm{HC}}$ with respect to the SOM used to infer systematic modes), or when trained on incomplete systematics information. Fig. \ref{kids:fig:t3_master} displays some performance examples of organised randoms, which we continue to explore in Appendix \ref{kids:app:datadriven_syst}. The panel titles in these figures give the SOM setups trained (i: bias) against KiDS-Bright to infer the clustering bias, and (ii: recovery) against the biased \flask mocks to recover the true clustering with organised randoms. Multiplicative factors in the panel titles indicate where $m\neq1$. Black points and dashed curves give the unbiased clustering signal $\pm1\sigma$ errors from \nflask \flask realisations (equivalent to red points in Fig. \ref{kids:fig:flask_wth}). Grey triangles and hatched curves are the measured (biased) clustering after depletion of the \flask fields, and solid filled curves show the corrected clustering, measured with organised randoms. In all meaningful (see Appendix \ref{kids:app:1point_correlations}, where we present a deliberate failure mode: \ttt{800Ares2}) bias and recovery cases we considered, our organised randoms yield clustering correlations that are more consistent with the truth than the biased signals (\ie those measured with uniform randoms).}

Fig. \ref{kids:fig:t3_master} shows that the recovery:\ttt{100C} organised randoms yield an effective correction of clustering biases from the bias:\ttt{800C} SOM (blue panel), with the recovered signal much closer to the unbiased measurement (mean absolute deviation over $7-100\,\rm{arcmin}$: $1.48\sigma\rightarrow0.31\sigma$, where $\sigma$ is the uncertainty in the unbiased signal). This indicates that a relatively insensitive SOM setup (\ttt{100C}) is able to characterise the more complex biases inferred from a SOM with greater sensitivity to small-scale systematic structure (\ttt{800C}). We see an even better correction ($1.43\sigma\rightarrow0.09\sigma$) when passing incomplete systematics information to the SOM, as demonstrated by bias:\ttt{800C} versus recovery:\ttt{100A} (orange panel). This demonstrates that our SOM methods are able to correctly infer systematic density fluctuation patterns even when the patterns are sourced by systematic-tracers that are unknown to the SOM. Inter-parameter correlations therefore serve to make organised randoms robust against missing training variables, and the additional freedom afforded to $N_{\rm{HC}}$ clusters in a systematics space of reduced dimensionality can improve the accuracy of the correction. 

\tbi{We note that these statements are specific to this particular setup. It is not generally true that SOMs will be able compensate for unknown systematics, especially if those unknowns levy significant density fluctuations and are uncorrelated with the variables presented in training of the SOM. Thus the selection of important systematic-tracers remains an intrinsic challenge for this type of study (one that can and should be tested with simulation-based methods similar to those presented here; also see \citealt{Weaverdyck2020}, who introduced methods for selecting the most important features from a large initial set). However, this setup does demonstrate that the elimination of redundant systematic-tracer features can improve the performance of clustering corrections by organised randoms (much as \citealt{Rezaie2019} demonstrated with their artificial neural network methods).}

Homogenising the scale sensitivity (through $N_{\rm{HC}}$) between bias and recovery SOMs, but keeping the incomplete systematics set for recovery (\ttt{100C} vs. \ttt{100A}; green panel), we still see an excellent recovery of the unbiased clustering signature ($0.94\sigma\rightarrow0.28\sigma$), as we might expect for a less complex bias. For identical bias and recovery setups \ttt{100C} (red panel), we begin to see a slight over-correction by the organised randoms. This comes about as clusters on the SOM begin to over-fit to the cosmic structure around the density-systematics relation.

\tbf{
Whilst the recovery is still preferable to the biased statistic here ($0.96\sigma\rightarrow0.54\sigma$), we acknowledge that over-corrections could be problematic for cosmological inference \citep[see][]{Wagoner2020,Weaverdyck2020}. However, the KiDS-Bright sample is a bright subset of the KiDS-$1000$ photometric sample, specifically chosen to be $\sim5$ magnitudes shallower than the survey flux limit, and thus is less sensitive to systematic detection failures; these data are already relatively unbiased. For more pathological biases, resulting in a higher amplitude of systematic density contrast, such over-fitting to cosmic structure is less likely to occur. Thus we expect our randoms to perform even better for samples with stronger systematics imprints, for instance the KiDS-$1000$ shear sample, dominated by faint galaxies. We tested this assertion using the same setup, but setting $m=4$ in Eq. \ref{kids:eq:Pdepletion}. In this case [\ttt{100C}$(\times4)$ versus \ttt{100C}; purple panel], a massively inflated bias of $12.17\sigma$ is once again reliably corrected to $0.34\sigma$.
}

\tbf{
\tr{This result is particularly important, as measuring accurate clustering statistics for faint, systematics-dominated samples has historically been extremely challenging; we therefore often choose bright subsamples for clustering analyses (also for typically more reliable estimation of the galaxy photo-$z$; see \citealt{Porredon2020}). Our companion letter, \tbf{Wright et al., (in prep.)}, thus presents an application of our organised randoms pipeline to the KiDS-$1000$ shear sample, finding excellent and robust performance in correcting for systematic clustering bias. Moreover, the shear sample is a factor of $\sim20$ denser on-sky than the bright sample, enabling us to increase the resolution of the cartesian RA/DEC grid whilst still respecting the survey mask (see Sec. \ref{kids:sec:artificial_syst_results}). In this way, we become more sensitive to small-scale systematic density fluctuations at fixed $N_{\rm{HC}}$, which \tbf{Wright et al., (in prep.)} show is important for these faint data.}
}

\section{Clustering in the KiDS-$1000$ bright sample}
\label{kids:sec:kids_clustering}

Having validated the performance of our SOMs and organised randoms in recognising systematic trends in galaxy density and removing their traces from synthetic galaxy two-point correlations, we applied our methods to measurements of galaxy clustering in the KiDS-Bright sample and compared tomographic cross-correlations with analogous signals measured in the highly complete GAMA \citep{Driver2009,Liske2015} sample.

We used randoms organised by the SOM configuration \ttt{100A} (detailed in Table \ref{kids:tab:SOMs}), which we show (in Figs. \ref{kids:fig:t3_master} \& \ref{kids:fig:t3_extra}) to robustly improve the fidelity of signal recovery in all bias:SOM scenarios from our \flask tests (which vary in scale sensitivity through $N_{\rm{HC}}$, and in the systematic-tracer parameters), thus we expect a reasonable correction even if the additional systematic-tracers (\ttt{MAG\_LIM\_r,EXTINCTION\_r,gaia\_nstar}) are in reality uncorrelated with the KiDS-Bright systematic density contrast. Moreover, the bias:\ttt{800C} / recovery:\ttt{100A} test case (Fig. \ref{kids:fig:t3_master}; orange panel) demonstrates that the bias inferred from all parameters (set \ttt{C}: threshold, PSF FWHM or ellipticity, limiting magnitude, Galactic density of stars and extinction), with high small-scale sensitivity ($N_{\rm{HC}}=800$), is corrected with great precision by \ttt{100A} organised randoms, having a residual bias of $<0.1\sigma$. Table \ref{kids:tab:test3_summary} summarises the performance of differently configured organised randoms and reveals \ttt{100A} to be the net-best performer across all bias:SOMs. 

We note that these particular SOM setups and parameter sets may not be ideal for other surveys with different areas, geometries, systematics imprints, etc., and that any organised randoms should be thoroughly tested with simulations, as we have done (in Sec. \ref{kids:sec:datadriven_systematics}).

\begin{figure*}[!htp]
    \centering
    \includegraphics[width=\linewidth]{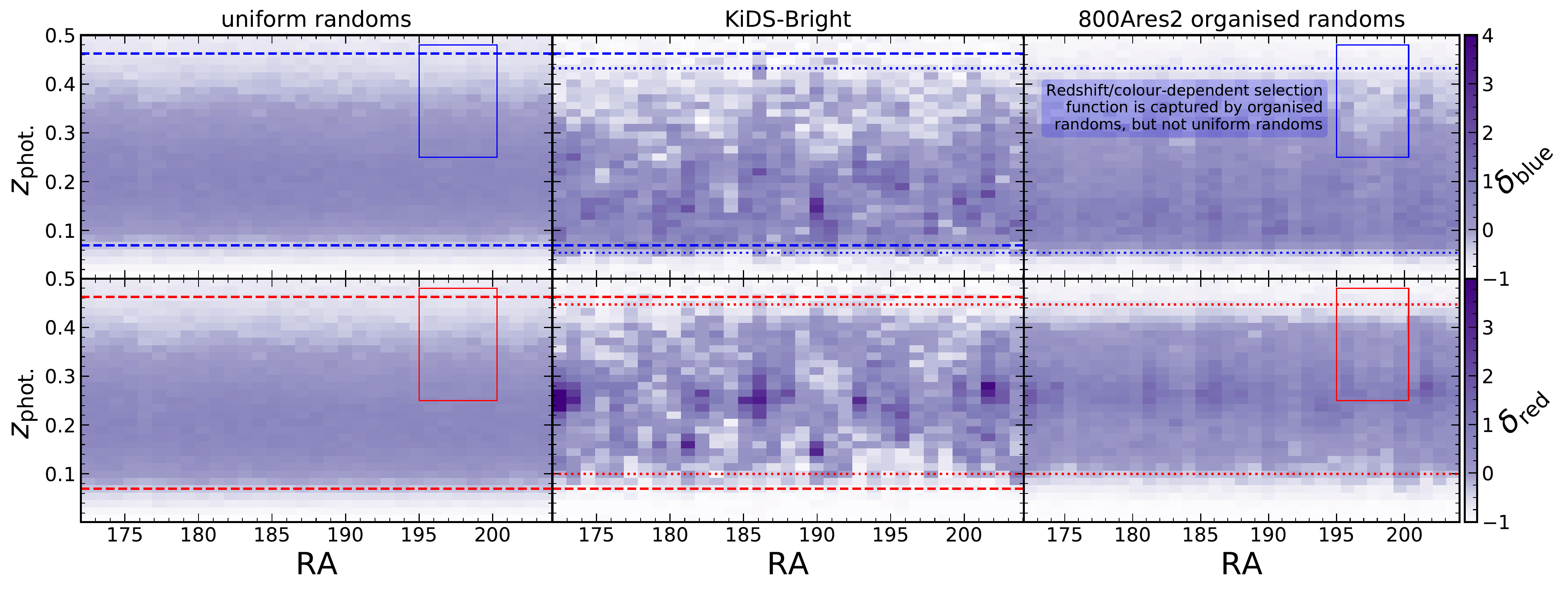}
    \caption[KiDS: demonstration of galaxy cloning utility]{Galaxy number density contrast $\delta$ in 2D bins of photometric redshift vs. RA for uniform randoms (\emph{left}), the KiDS-$1000$ bright sample (KiDS-Bright; \emph{middle}), and relevant clones from \ttt{800Ares2} organised randoms (\emph{right}). Rows display blue (\emph{top}) and red (\emph{bottom}) galaxies, according to a boundary at observer-frame $u-i=2.66$. We only show galaxies in the range $172<\rm{RA}<205$, $2<\rm{DEC}<3$ in order to reveal systematic density variations in the data and randoms. Unlike for uniform randoms, the organised randoms can be seen to reproduce the systematic trends in the KiDS-Bright data, namely under-densities over different pointings (\eg ${\rm{RA}} = 175.5,197.5$), visible as vertical bands of fainter pixels, and the differential evolution of densities in the samples with redshift, as evidenced by the boxes (highlighting a specific volume with differential density as a function of sample colour) and horizontal lines (dashed lines give the $z_{\rm{phot.}}$ corresponding to the $3rd^{\rm{}}$ and $97th^{\rm{}}$ percentile number counts for uniform randoms, and dotted lines give these for KiDS-Bright data). We elect to display \ttt{800Ares2} randoms here for a clearer illustration of the cloning mechanism, which is more subtle in our other favoured randoms (see Sec. \ref{kids:sec:test3_results}). 
    }
    \label{kids:fig:ravsz}
\end{figure*}

Fig. \ref{kids:fig:ravsz} indicates the advantages of creating cloned galaxy randoms in which each random point is a clone of a real object used in training of the SOM, and clones are spatially restricted to on-sky areas occupying similar positions in the systematics space. The figure illustrates how we reproduced systematic density variations in a strip of KiDS-Bright ($172<\rm{RA}<205$, $2<\rm{DEC}<3$ degrees) for a coarse red and blue (bottom and top) sample selection\footnote{We defined the red and blue boundary at observer-frame $u-i=2.66$, which sits in the trough of the bimodal colour distribution of KiDS-Bright.} merely by restricting our organised randoms to the relevant clones. The figure shows the density contrast of 1 degree columns in RA, and we expand the radial dimension ($z_{\rm{phot.}}$) to draw attention to non-physical density modes, that is, underdensities that are localised to single pointings, seen as fainter vertical strips of colour in the right-most column (\eg at ${\rm{RA}}\in[\,175.5,197.5\,]$), and to the redshift evolution of galaxy number density, which varies for red and blue galaxies; these trends are mirrored in the \ttt{800Ares2}\footnote{Chosen for illustrative purposes, as the inferred $\delta_{\rm{OR}}$ is wider for this setup \hlstop{(}see Appendix \ref{kids:sec:test2_results}\hlstop{)} and the increased resolution results in more clearly visible systematic density modes in the organised randoms (discussed in Appendices \ref{kids:app:1point_correlations} and \ref{kids:app:datadriven_syst}).} organised randoms (right column) in contrast with the uniform randoms (left column). The cloning utility will be fully realised in future work. Systematics that differentially affect arbitrary galaxy selections can be automatically compensated for, allowing for easy splitting of analyses into red and blue galaxies, for instance, or into bins of galaxy luminosity. For now, we focus only on redshift tomography.

In addition to the tomographic bins 1 and 2, with edges at $z_{\rm{phot.}}=\{0.02,0.2,\text{and }0.5\}$, we cross-correlated two additional, non-overlapping bins, 1a and 2a, with edges at $z_{\rm{phot.}}=\{0.02 \text{ and }0.22\}$ and $z_{\rm{phot.}}=\{0.28   \text{ and }0.5\}$, respectively. The gap between the inner edges of the two bins is more than the typical 95\% photo-$z$ scatter at this redshift (Fig. \ref{kids:fig:flask_wth}, red line in the top panel), therefore we tested for a null cross-correlation between them. Barring the auto-correlations of the full (total) sample, we refer to correlations between bins $i$ and $j$ as $i$-$j$, that is, 1-1 is the auto-correlation of tomographic bin 1 and 1-2 is the cross-correlation between bins 1 and 2, etc.

\begin{figure}
    \centering
    \includegraphics[width=\columnwidth]{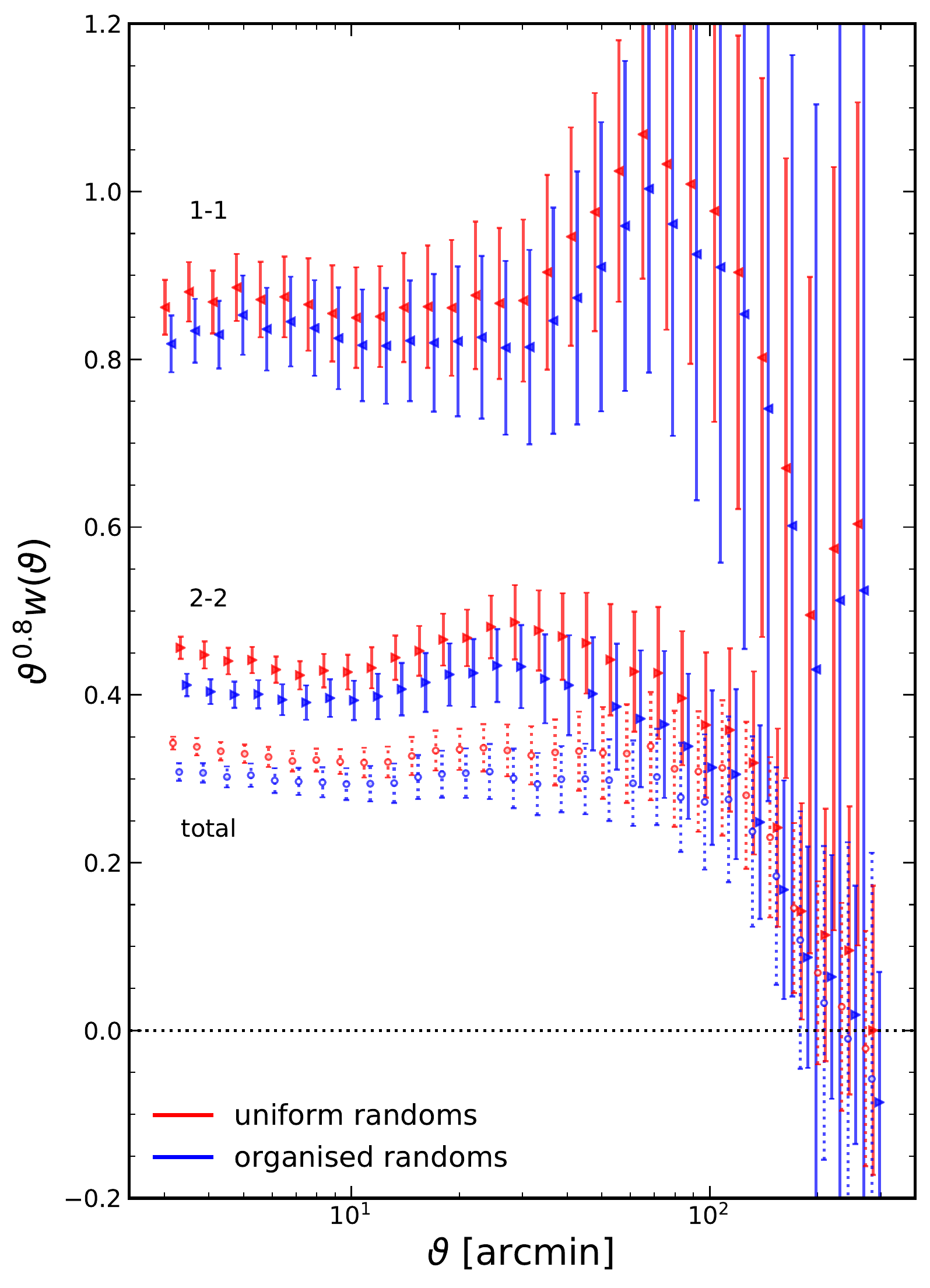}
    \caption[KiDS: \wth auto-correlation signals measured in KiDS-$1000$ bright sample (KiDS-Bright), with/out corrective randoms]{Angular auto-correlation functions \wth measured in the total (\ie unselected) KiDS-$1000$ bright sample (\emph{bottom}), and within redshift bins 1 (1-1; \emph{top}) and 2 (2-2; \emph{middle}), with edges $\in[\,0.02,0.2,0.5\,]$, as shown in Fig. \ref{kids:fig:flask_wth}. Measurements using uniform randoms are shown in red, and those made with \ttt{100A} organised randoms are shown in blue. Errors are estimated with a 2D delete-one jackknife with 31 pseudo-independent patches of the footprint. Points are horizontally offset to aid clarity. The total sample correlation is more clearly visible in Appendix Fig. \ref{kids:fig:gama_comparison}.
    }
    \label{kids:fig:k1000clustering}
\end{figure}

\begin{figure}
    \centering
    \includegraphics[width=\columnwidth]{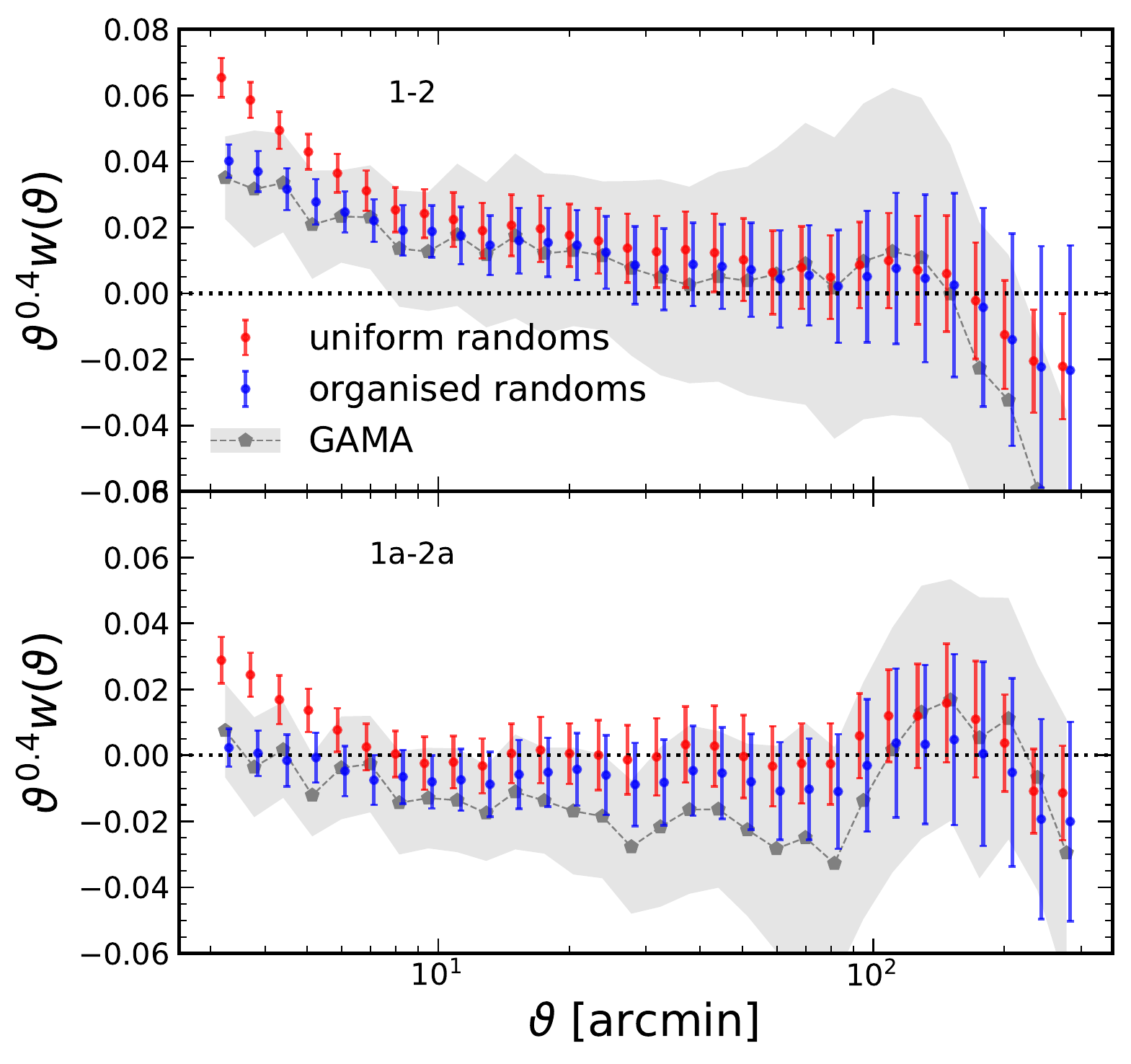}
    \caption[KiDS: \wth cross-correlation signals measured in KiDS-Bright, with/out corrective randoms]{Angular cross-correlation functions \wth measured between KiDS-$1000$ bright sample (KiDS-Bright) redshift bins 1 and 2 (\emph{top}), with edges $\in[\,0.02,0.2, \text{and }0.5\,]$, and 1a and 2a (\emph{bottom}), with edges $\in[\,0.02 \text{ and }0.22\,]$ and $[\,0.28    \text{ and }0.5\,]$, as shown in Fig. \ref{kids:fig:flask_wth}. Measurements using uniform randoms are shown in red, and those made with \ttt{100A} organised randoms are shown in blue. Errors are estimated with a 2D delete-one jackknife with 31 pseudo-independent patches of the footprint. Grey points and shading give the equivalently binned (by {\sc{ANNz2}} photo-$z$) correlations measured for the GAMA sample, with errors estimated again with a jackknife, but from 20 sub-regions of the GAMA window. Errors on scales $\vartheta\gtrsim180\,\rm{arcmin}$ are thus likely to be underestimated for GAMA correlations.
    }
    \label{kids:fig:k1000clustering_cc}
\end{figure}

We generated organised randoms with $20   \text{ times}$ the number density of KiDS-Bright, so as to combat Poisson-noise in the relevant pair-counts (see Eq. \ref{kids:eq:wth_landyszalay}). Our measurements of angular clustering auto-correlations within the two primary redshift bins and across the full KiDS-Bright sample are displayed in Fig. \ref{kids:fig:k1000clustering}. Fig. \ref{kids:fig:k1000clustering_cc} displays the redshift bin cross-correlations  in comparison with equivalently binned signals measured in the GAMA sample. We used our {\sc{ANNz2}} photo-$z$ estimates to define these GAMA bins and employed the spatially uniform, windowed, cloned-galaxy randoms presented by \cite{Farrow2015}. As we discuss in Appendix \ref{kids:app:gama_comparison}, auto-correlations in GAMA are not suitable for validating KiDS-Bright correlations here, hence they are not displayed in Fig. \ref{kids:fig:k1000clustering}.

For all \wth correlations in Figs. \ref{kids:fig:k1000clustering} and \ref{kids:fig:k1000clustering_cc}, we estimated errors with the delete-one jackknife technique. We divided the KiDS-Bright/GAMA footprints into $N_{\rm{patch}}$ roughly equal-area patches, and computed \wth upon the successive removal of individual patches. For these $N_{\rm{patch}}$ signals $w_{\alpha}$, the covariance is then

\begin{equation}
    \hat{C}_{\rm{jack.}} = \frac{N_{\rm{patch}}-1}{N_{\rm{patch}}} \, \sum_{\alpha=1}^{N_{\rm{patch}}} \, \, (w_{\alpha} - \bar{w})(w_{\alpha} - \bar{w})^{\rm{T}} \quad ,
    \label{kids:eq:covariance}
\end{equation}
where $\rm{T}$ denotes the conjugate transpose of a vector and $\bar{w}$ is the average of the $N_{\rm{patch}}$ measurements $w_{\alpha}$. For KiDS-Bright, $N_{\rm{patch}}=31$, and $N_{\rm{patch}}=20$ for GAMA\footnote{Since the total equatorial GAMA area $\sim180\sqdeg$, each of the 20 patches is $\sim9\sqdeg \equiv [180\,\rm{arcmin}]^{2}$ in size, hence errors on $w(\vartheta\gtrsim180\,\rm{arcmin})$ are likely to be under-estimated.}. We note that for our chosen binning of 30 log-spaced bins in the range $3\leq\vartheta\leq300\,\rm{arcmin,}$ so few jackknife samples will yield singular covariance matrices \citep[see][]{Hartlap2007}. We are only interested in measurement errors (the square root of the matrix diagonal) here, and because we did not perform any fitting, we proceeded accordingly.

Fig. \ref{kids:fig:k1000clustering} shows that the \ttt{100A} organised randoms make more sizeable corrections (blue points) to the measured total and 2-2 correlations (red points) than to the 1-1 correlation. This is perhaps to be expected because\tbf{ at higher redshifts the objects with faint apparent magnitudes, and predominantly small angular extents, will be more prone to dipping below the detection threshold (or out of sample selection criteria) as a result of observational effects; the smaller correction to 1-1 could then be the opposite manifestation of this effect, wherein apparently brighter or larger galaxies are more robustly detected.} The large-scale corrections to cross-correlations (Fig. \ref{kids:fig:k1000clustering_cc}) are also small. This is the desired behaviour, particularly in the case of disjoint bins 1a-2a, as the cross-correlation is expected to be zero. The rising 1a-2a signals at low values of $\vartheta$ (red points) are largely corrected for by the organised randoms. Given the vast gap ($\delta{z}=0.06$) between the disjoint bins, these small-$\vartheta$ signals are highly unlikely to correspond to real structures. The organised randoms \hlstop{thus} compensate for small-scale systematic density modes that are shared across the redshift range. 

We note the (still relatively small) amplitude of the \wth correction at $\vartheta\gtrsim10\,\rm{arcmin}$ in the total sample correlation (in Fig. \ref{kids:fig:k1000clustering}) and its similarity to the signal recoveries displayed in Fig. \ref{kids:fig:t3_master}. This is encouraging as we can intuit that our extraction and replication of depletion patterns, from KiDS-Bright and in \flask realisations (Sec. \ref{kids:sec:flask_biasing}), is realistic enough to result in consistent inferences of the required correction to $w(\vartheta)$. Thus we can expect that the corrected total signal is closer to the true clustering, as is the case for the correlations from Fig. \ref{kids:fig:t3_master}. In the intermediate range $7<\vartheta<100\,\rm{arcmin}$, the mean corrections to \wth for each correlation are \{1-1: $-5.6\%$, 1-2: $-30.0\%$, 2-2: $-10.9\%$, and total: $-9.4\%$\}.

In order that analyses be conducted free of any confirmation bias, it has become common in large-scale structure analyses to practice \emph{\textup{blinding}} \citep[\eg][]{Kuijken2015}, wherein several modified versions of key statistics are produced\footnote{The best methods for blinding are active areas of research; see for example \cite{Muir2019}, \cite{Sellentin2020}, and \cite{Brieden2020}.} from the data alongside the true version. The truth is revealed to the team by an independent entity only after the entire analysis is complete, such that no critical decisions can be taken in favour of some expected result. At the time of writing, we were blind to the truth of our KiDS data, and thus chose not to compare with theoretical models for galaxy clustering. \tr{Our companion work (Wright et al., in prep.) considers the best-fit cosmological model from the latest KiDS 3x2-point analysis \citep{Heymans2020} for comparison with corrected clustering statistics measured in the KiDS-$1000$ shear sample.
}

A direct comparison of GAMA clustering auto-correlations with our own is likewise unsuitable here for reasons we discuss in Appendix \ref{kids:app:gama_comparison}. We therefore only considered the redshift bin cross-correlations from GAMA (in Fig. \ref{kids:fig:k1000clustering_cc}) for validation. With the alignment of sample properties less important here, these correlations instead probe (i) the photo-$z$ scatter and (ii) any systematic correlations shared across the KiDS-Bright redshift bins, where the former should be negligible in the 1a-2a correlation owing to the large gap (Fig. \ref{kids:fig:flask_wth}) between the bins. The rising signals at small-$\vartheta$ (Fig. \ref{kids:fig:k1000clustering_cc}; red points) are absent from GAMA (grey), and our organised randoms' corrections (blue) result in greater consistency between GAMA and KiDS-Bright. This is also true of the slightly negative (but not significant) 1a-2a blue data-points in the $\vartheta\sim10\,\rm{arcmin}$ range, where the negative GAMA points indicate possible LSS fluctuations. The GAMA data are highly complete ($>98\%$; \citealt{Liske2015}) and can be considered as the unbiased truth here. They are what we expect for photo-$z$ of this quality in the absence of systematic galaxy density patterns. We thus argue that our corrections to the redshift bin cross-correlations successfully remove systematic correlations from KiDS-Bright data.

\section{Summary}
\label{kids:sec:discussion}

We have developed and tested a method for the construction of organised randoms, which mirror systematic trends in galaxy density, using self-organising maps. We made extensive use of lognormal random field simulations from \flask to test the abilities of SOMs to recognise both artificial and real systematic loss of galaxies, and demonstrated that organised randoms constructed using this information are able to reliably correct the measurable angular clustering of the synthetic data. 

With the present data volume, constructing effective organised randoms relies upon a balance between the area of sky probed by each hierarchical cluster (essentially an $n$-dimensional bin) defined on the SOM, the variables and dimensionality of the systematics space given to the SOM, and the width of the distribution of \emph{\textup{systematic}} density contrast. As we have demonstrated, this balance is readily assessed with simulations. If systematic modes are very much smaller than cosmological modes, organised randoms become more prone to over-correction of clustering biases, although our SOM methods are able to test for the necessity of any correction, as they estimate the distribution of systematic density contrast $\delta_{\rm{OR}}$. Conversely, for strong pathological density modes, our randoms are highly effective. Moreover, regardless of our analysis choices when testing with \flask simulations, our recovered clustering results are always consistent with the underlying truth, having an average bias correction across all meaningful runs of $2.31\sigma\rightarrow0.34\sigma$. Our recovery of the truth is particularly striking in the amplified bias ($m=4$) case, where we shift from catastrophic bias with uniform randoms ($12.2\sigma$) to full consistency with the truth ($0.34\sigma$).

We found that the importance of certain systematics-tracing variables at the level of the two-point correlation function is not necessarily determined by the strength of the one-point (pixel) correlation of the tracer with galaxy density (see Appendix \ref{kids:app:1point_correlations}). Whilst this finding may only hold for the KiDS-$1000$ bright sample (KiDS-Bright), we note that simply correcting for these one-point correlations may yet be problematic for the general goal of recovering unbiased galaxy clustering two-point functions. The amount of one-point correction required to achieve this goal is not necessarily clear and is further complicated by correlations between systematic-tracer parameters. We recommend the use of principle component analysis (PCA) to alleviate the latter concern, but two-point functions should also be considered for validation in analyses of this type.

\tbf{We worked with \flask simulations, modelling the footprint and number density of the KiDS-$1000$ bright sample, to create realistically biased synthetic galaxy fields within which to test the performance of our organised randoms. We inferred the field of systematic galaxy density contrast directly from KiDS-$1000$ bright sample data under various assumptions modifying sensitivity to angular scales and to different systematic-tracers, and modifying the amplitude of systematic fluctuations. Applying these data-driven systematic clustering imprints to many independent realisations from {\sc{flask}}, we then generated organised randoms by training against the biased \flask fields, again varying the scales and tracers we used. Under several scenarios of biasing due to the spatially variable PSF, detection threshold, survey depth, and Galactic stellar density and extinction, we found that training upon the \ttt{psf\_fwhm,psf\_ell,MU\_THRESHOLD} parameters and defining 100 hierarchical clusters on the trained SOM (setup \ttt{100A}) was sufficient to yield organised randoms that consistently remove the various realistic systematic density modes from the \flask galaxy fields. These \ttt{100A} organised randoms recovered clustering signals deviating on average from the unbiased signal at $\sim0.3\sigma$ over the relevant {\sc{flask}}-testing setups with average bias $\sim1.1\sigma$. For the most pessimistic clustering bias scenarios, where uncorrected signals deviate from the truth at up to $\sim12\sigma$, the performance of organised randoms remains robust, with the bias of recovery at $\sim0.3\sigma$.}

We presented the first measurement of photometric galaxy clustering from the KiDS for bright GAMA-like galaxies from the $1000\sqdeg$ $4th^{\rm{}}$ Data Release. Defining two tomographic bins with edges $z_{\rm{phot.}} = \{0.02,0.2,\text{and }0.5\}$, we measured the angular auto- and cross-correlation functions over $3<\vartheta<300\,\rm{arcmin}$ with uniform and organised randoms. We saw that our organised randoms make variable corrections to tomographic auto- and cross-correlations, editing amplitudes at intermediate angular scales ($7\lesssim\vartheta\lesssim100\,\rm{arcmin}$) by up to $\sim10\%$ ($\sim1\sigma$) in the auto-correlations, and $\sim30\%$ in the cross-correlations.

\tbf{We implemented our randoms such that each random point was a clone of a real galaxy, scattered within regions of the survey footprint that are similar to the location of the parent galaxy in terms of the position it occupies in systematics space. Thus by mimicking galaxy sample selections in the randoms, we compensated for distinct sample-specific systematic correlations such as those induced by selections in galaxy photo-$z$. For tomographic cross-correlations, our randoms were found to correct significant systematic density modes at small-$\vartheta$, which are shared between disparate redshift populations, whilst making nearly negligible corrections throughout the remaining angular range. This indicates similar small-scale, but distinct larger-scale systematic clustering imprints for the different redshift populations. This utility is easily generalised to any galaxy sample selections in luminosity, colour, etc.}

An extension of this work to increased areas and galaxy number densities is extremely promising. Larger areas will result in better handling of large-$\vartheta$ systematic density modes due to more redundant sampling and in a smoother distribution of density contrast, which will minimise contamination of the randoms by cosmic structure. Higher galaxy densities offer better sampling of small-$\vartheta$ modes, a smoother description of the systematics space, and the \hlstop{possibility} of increasing the resolution of the randoms without fear of greater contamination by structure. Thus the performance of organised randoms should improve on all scales with next-generation datasets; \tr{our companion letter \tbf{(Wright et al., in prep.)} moves to verify our assertions here, applying our testing pipeline to measurements of clustering in the faint KiDS-$1000$ shear sample, thus exploring a deep-survey, high number density scenario.}

Upon \hlstop{publication of our companion letter}, we will make our code and methods public, such that independent teams can experiment with the handling of systematic density variations using organised random clones. Future surveys that will make powerful use of galaxy clustering (\eg the Rubin Observatory and \emph{Euclid}) can then include the construction of organised randoms as a pipeline module, adding to the panoply of complementary means for the accurate measurement of galaxy positional statistics.

{\tiny
\textbf{Acknowledgements:} We thank Chris Morrison and Boris Leistedt for helpful discussions during the early phase of this work. HJ acknowledges support from a UK Science \& Technology Facilities Council (STFC) Studentship. This work is part of the Delta ITP consortium, a program of the Netherlands Organisation for Scientific Research (NWO) that is funded by the Dutch Ministry of Education, Culture and Science (OCW). AHW, AD, HHi are supported by a European Research Council Consolidator Grant (No. 770935). MB is supported by the Polish Ministry of Science and Higher Education through grant DIR/WK/2018/12, and by the Polish National Science Center through grants no. 2018/30/E/ST9/00698 and 2018/31/G/ST9/03388. BG acknowledges support from the European Research Council under grant number 647112 and from the Royal Society through an Enhancement Award (RGF/EA/181006). CH acknowledges support from the European Research Council under grant number 647112, and support from the Max Planck Society and the Alexander von Humboldt Foundation in the framework of the Max Planck-Humboldt Research Award endowed by the Federal Ministry of Education and Research. HHi is supported by a Heisenberg grant of the Deutsche Forschungsgemeinschaft (Hi 1495/5-1). HHo acknowledges support from Vici grant 639.043.512, financed by the Netherlands Organisation for Scientific Research (NWO). MV acknowledges support from the Netherlands Organisation for Scientific Research (NWO) through grant 639.043.512. This work was initiated at the Aspen Center for Physics, which is supported by National Science Foundation grant PHY-1607611.

This work is based on observations made with ESO Telescopes at the La Silla Paranal Observatory under programme IDs 100.A-0613, 102.A-0047, 179.A-2004, 177.A-3016, 177.A-3017, 177.A-3018, 298.A-5015, and on data products produced by the KiDS consortium. GAMA is a joint European-Australasian project based around a spectroscopic campaign using the Anglo-Australian Telescope. Our GAMA catalogue is based on data taken from the Sloan Digital Sky Survey and the UKIRT Infrared Deep Sky Survey. Complementary imaging of the GAMA regions is being obtained by a number of independent survey programmes including GALEX MIS, VST KiDS, VISTA VIKING, WISE, Herschel-ATLAS, GMRT and ASKAP providing UV to radio coverage. GAMA is funded by the STFC (UK), the ARC (Australia), the AAO, and the participating institutions. The GAMA website is \url{http://www.gama-survey.org/}.

Some of the results in this paper have been derived using the healpy and {\sc{HEALPix}} packages.

\emph{Author contributions:} All authors contributed to the development and writing of this paper. The authorship list is given in two groups: the lead authors (HJ, AHW, BJ), followed in alphabetical order by authors who made a significant contribution to either the data products or to the scientific analysis.
}

\bibliographystyle{aa}
\bibliography{RandomsRefs}

\appendix

\section{Artificial systematics}
\label{kids:app:test1}

\begin{figure*}
    \centering
    \includegraphics[width=\linewidth]{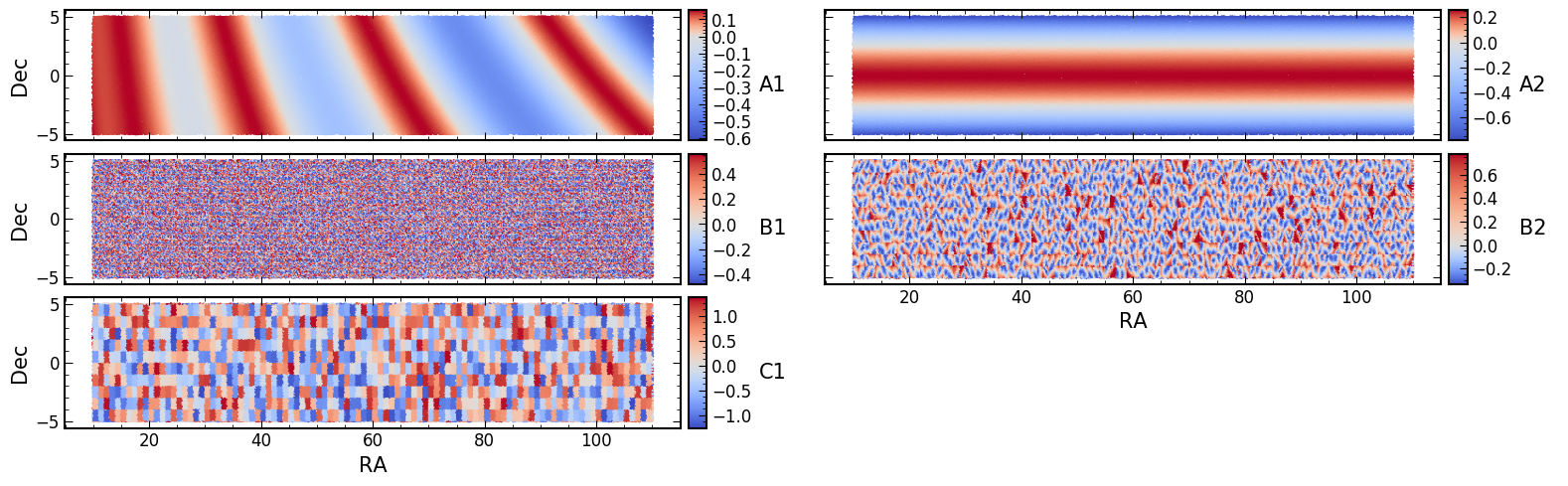}
    \caption[{\sc{flask}}: maps of the excess probability of depletion as a function of artificial systematics-variables]{Same as Fig. \ref{kids:fig:imagpars}, but only for parameters with non-zero depletion functions. The colouring now denotes the \emph{\textup{excess}} probability of depletion as a function of each artificial systematic. A galaxy with an excess probability of 1 is 100\% more (or twice as) likely to be depleted when compared with an excess probability of zero.
    }
    \label{kids:fig:excess_deplprob}
\end{figure*}

\begin{figure*}[!ht]
    \centering
    \includegraphics[width=\linewidth]{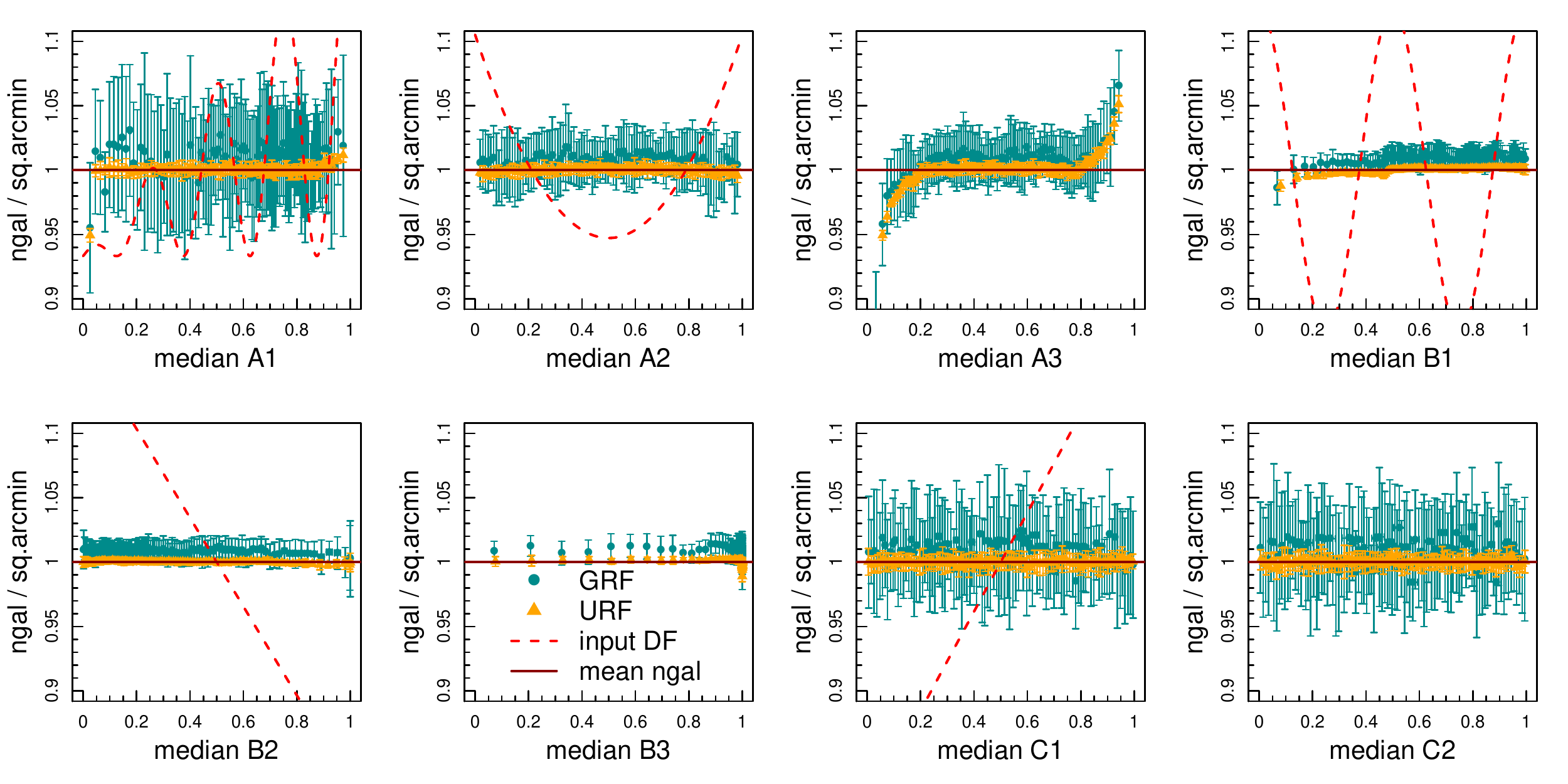}
    \caption[{\sc{flask}}: the self-organising map's inferred null depletion functions in unbiased lognormal/uniform random fields]{Same as Fig. \ref{kids:fig:test1_tob}, but for a SOM trained on an unbiased \flask field. The depletion functions are again shown as dashed red lines, although no depletions were applied. Irregularities in the \ttt{A3} panel, for example, are discussed in Sec. \ref{kids:sec:artificial_syst_results}. Data-points derived from lognormal random fields (green) are on average higher in $n_{\rm{gal}}$ than those from uniform fields (orange) because of the clustering of galaxies. The Cartesian grid we used to estimate the area of the sample has more empty cells for the GRF, hence the area is under-estimated and $n_{\rm{gal}}$ is over-reported.
    }
    \label{kids:fig:test1_tou}
\end{figure*}

Here we extend our discussion of the characterisation of artificial systematic density fluctuations with self-organising maps (Sec. \ref{kids:sec:artificial_systematics}). We created artificial systematic-tracer variables (Fig. \ref{kids:fig:imagpars}) with arbitrary depletion functions (Fig. \ref{kids:fig:depletion_functions}) to apply to \flask mock data (the excess probability of depletion for each artificial variable is given in Fig. \ref{kids:fig:excess_deplprob}), finding the \ttt{T1mock} SOM (Table \ref{kids:tab:SOMs}) to be capable of characterising the depletion functions directly from the mock catalogue (Fig. \ref{kids:fig:test1_tob}). As a sanity check, we also ran the same SOM against the \nflask \flask catalogues \emph{\textup{before}} applying any depletion. The SOM should not find any density-systematic correlations where none exist. The results are shown in Fig. \ref{kids:fig:test1_tou}, where the depletion functions from Fig. \ref{kids:fig:depletion_functions} are included for reference, and we see the desired behaviour. The SOM recognises no significant, systematic trends in galaxy density (barring some of the irregularities related to grid resolution, discussed in Sec. \ref{kids:sec:artificial_syst_results}).

\section{Pixel density systematic one-point correlations}
\label{kids:app:1point_correlations}

Here we consider correlations between the densities of galaxies in pixels on-sky and the values of systematic-tracer variables in those pixels. Many versions of these statistics appear in the literature, often as metrics to demonstrate successful decorrelation of galaxy densities and systematics by other methods \citep[\eg][\tbi{where some also use two-point metrics, which we argue is necessary}]{Suchyta2016,Rezaie2019,Kitanidis2019} and occasionally in order to \tbi{directly} derive the corrections themselves \citep{Elvin-Poole2017,Vakili2020}. \tbi{We note that the drawbacks of one-point metrics that we explore in detail here may be specific to our methods and the parameters of our survey data.} For these tests, we constructed galaxy density and systematics maps using \ttt{healpy} with $n_{\rm{side}}=256$, corresponding to on-sky pixels of roughly $14\times 14\,\rm{arcmin}$ in size. We computed the mean systematics in each pixel, applying requisite scaling to pixels that were only partially sampled by the survey window. \tbi{Taking the means-in-pixels will constitute some loss of information, but this should primarily affect sub-pixel scales. We leave any exploration of higher resolution maps and/or improved pixel statistics to future work with denser samples.}

This approach is somewhat analogous to that of \cite{Elvin-Poole2017}, who used \tbi{individual density systematic} one-point functions \tbi{directly} to derive per-galaxy weights for the mitigation of clustering systematics. In our case, however, the one-point correlations serve only as a diagnostic. Our SOM approach improves upon \tbi{direct} pixel-correlation methods and drops some commonly made approximations. In particular, we mapped the galaxy depletion function through a higher-dimensional systematics space, correcting galaxies directly in this space, rather than iteratively suppressing only the worst \tbi{individual} correlations until crossing some threshold. \tbi{We note that \cite{Weaverdyck2020} showed with DES year-one-like simulations that the latter point does not damage the success of corrections}. Additionally, whilst usually approximated as linear in the density contrast, the depletion of galaxies as a \tbi{multivariate} function of systematics in our method can be arbitrarily non-linear, allowing greater freedom to correct for more complex systematic modes

\hlstop{We note that, }similarly to this analysis, \cite{Rezaie2019} tested their mitigation scheme with contaminated mock catalogues, finding that linear models for density-systematic correlations were sufficiently descriptive to yield successful corrections. However, their contamination of mocks was itself based upon the linear model, thus the necessity of non-linear corrections was not fully explored for those data. We introduced methods for contamination of mocks based upon the non-linear density-systematics modelling from SOMs (Sec. \ref{kids:sec:flask_biasing}). Future work can apply linearly derived corrections to these mocks and test for the necessity of non-linearity in the modelling for any dataset.

We made some changes to the typical formulation of the galaxy density versus systematic pixel correlations. First, we normalised per-pixel number counts (or pixel counts) of galaxies, $N_{\rm{gal}}$, not by the global average galaxy pixel counts, $\left\langle{}N_{\rm{gal}}\right\rangle$, but instead by pixel counts for a uniform random field, $N_{\rm{rand}}$ (given the same on-sky pixelisation). \tbi{These considerations are particularly relevant for our method, which constructs a high-resolution mask of the sky using data and random points.} The two estimators are generally equivalent in the limit of large area, modulo normalisation between the average random and galaxy number densities. Second, for the corrected case, we normalised the galaxy pixel counts by the organised randoms' pixel counts. Any systematic correlations in the data should thus be cancelled by the mirrored trends in the organised randoms. 

We note here, however, that the erasure of such trends can be a misleading measure of performance. Inter-tracer correlations can induce trends in the $N_{\rm{gal}}\,/\,N_{\rm{rand}}$ distribution as a function of systematics that make individual systematics appear more or less significant than they truly are \tbi{(sometimes referred to as `chance correlations' in the literature)}. Furthermore, the strength of correlation per systematic that is needed to introduce a significant clustering bias is also an open question, but is likely dependent upon the sample being analysed, the distribution and dynamic range of the systematic parameter, and the ultimate angular scales of interest. Finally, these distributions can behave particularly pathologically when correcting systematic density variations using variable randoms, as we did here. As a simple demonstration, should the organised randoms produce a catalogue that exactly reproduced the input galaxy catalogue (\ie the randoms encoded all galaxy density fluctuations, both systematic and cosmological), then these correlations would appear perfectly flat. This would constitute the most abject failure of our corrective method, and yet this diagnostic would be labelled a success. 

These concerns led us to derive and consider orthogonal principle component maps \citep[\ie principle component analysis, or PCA; see][for a comprehensive text on PCA]{jolliffe2002principal} alongside our systematics maps \citep[see also][who applied such a decomposition to their systematics maps]{Wagoner2020}. Principle components are defined as new orthogonal basis vectors for $n$-dimensional data, constructed from linear combinations of the original dimensions. The new basis then describes in order, from component 1 to component $n,$  the directions of greatest variance in the data. As the basis vectors are orthogonal, principle components have no covariance and are thus more instructive for assessing the significance of density-systematic correlations, although they are more difficult to interpret physically.

\begin{figure*}
    \centering
    \includegraphics[width=\textwidth]{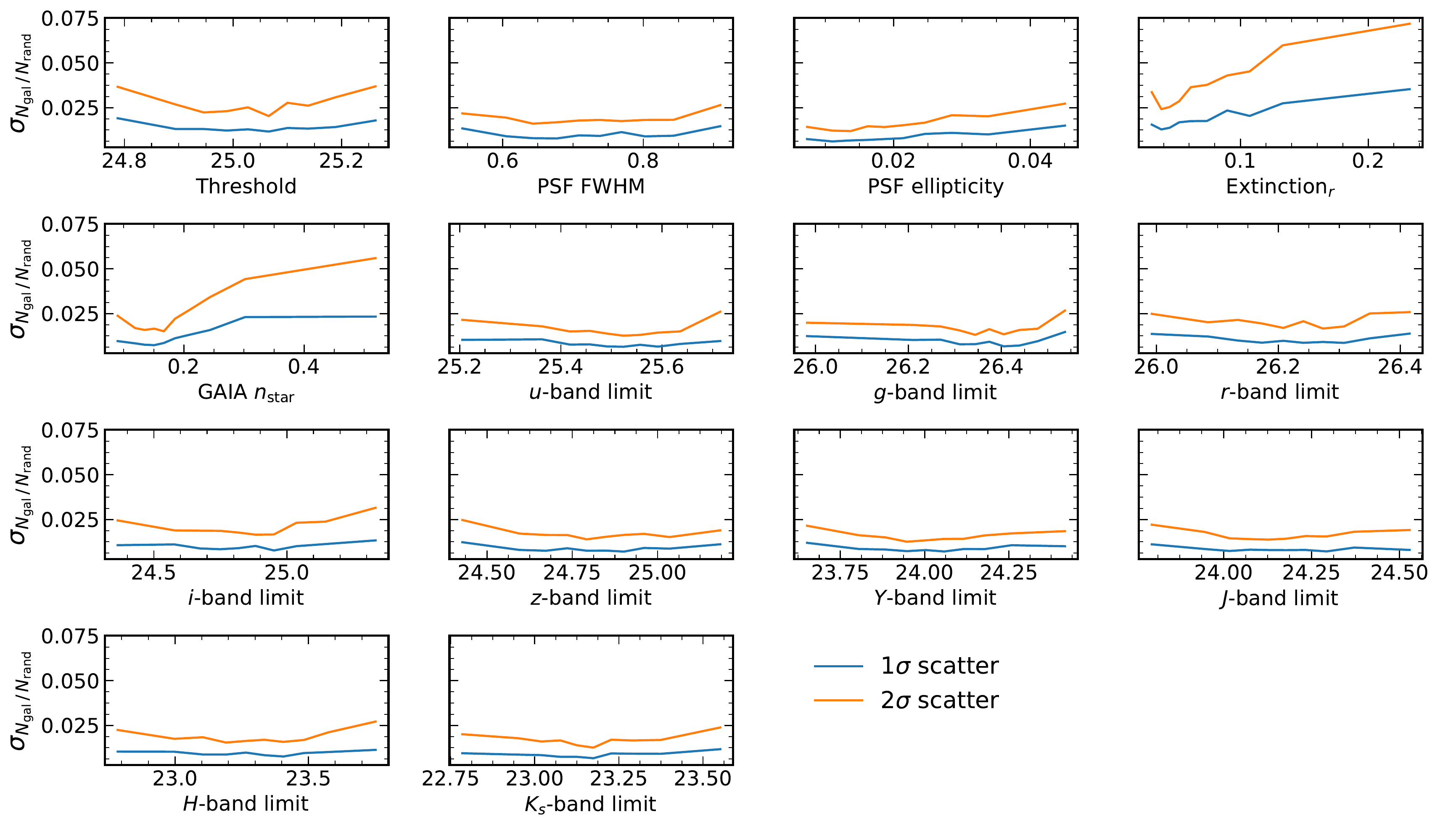}
    \caption[KiDS and {\sc{flask}}: the intrinsic $1\sigma$ and $2\sigma$ spread of pixel correlations between galaxy number density in 100 \flask fields and systematic-tracer variables interpolated onto \flask galaxies from the KiDS bright sample]{Intrinsic scatter of $N_{\rm{gal}}\,/\,N_{\rm{rand}}$ vs. various systematics tracer variables (Table \ref{kids:tab:housekeeping_variables}) as measured over 100 \flask realisations with the KiDS-Bright survey parameters or footprint. Without any biases applied to the \flask realisations, the blue ($1\sigma$) curves give half the difference between the $16th^{\rm{}}$ and $84th^{\rm{}}$ percentiles of the distribution of 100 individual $N_{\rm{gal}}\,/\,N_{\rm{rand}}$ vs. systematics correlations. $2\sigma$ (orange) curves give half the difference between the $2.5th^{\rm{}}$ and $97.5th^{\rm{}}$ percentiles. Galaxy/random counts and mean systematics are taken in $n_{\rm{side}}=256$ pixels, about $14\,\rm{arcmin}$ in size.}
    \label{kids:fig:expected_scatter_t3}
\end{figure*}

Lastly, we assessed the expected spread of pixel density-PCA component (or density-systematic) correlations for individual unbiased realisations of the large-scale structure within our survey parameters. These are given in Fig. \ref{kids:fig:expected_scatter_t3}, where solid lines give the $1\sigma$ and $2\sigma$ widths of the $N_{\rm{gal}}\,/\,N_{\rm{rand}}$ versus systematic-tracer distributions from 100 individual \flask realisations. As Fig. \ref{kids:fig:expected_scatter_t3} shows, the intrinsic spread of galaxy number density versus systematic-tracers over 100 independent realisations of KiDS-Bright-like synthetic data is often around a few percent and rises for more long-varying tracers such as Galactic extinction and stellar density. Thus, for our data, galaxy density-component or systematic correlations within a few percent of unity cannot be distinguished from those arising stochastically. These are the limits of useful bias-correction dictated by sample variance, and we mark them as purple shading in the following pixel-correlation figures.

\subsection{Correction of one-point correlations}
\label{kids:sec:test2_results}

\begin{figure}
    \centering
    \includegraphics[width=\columnwidth]{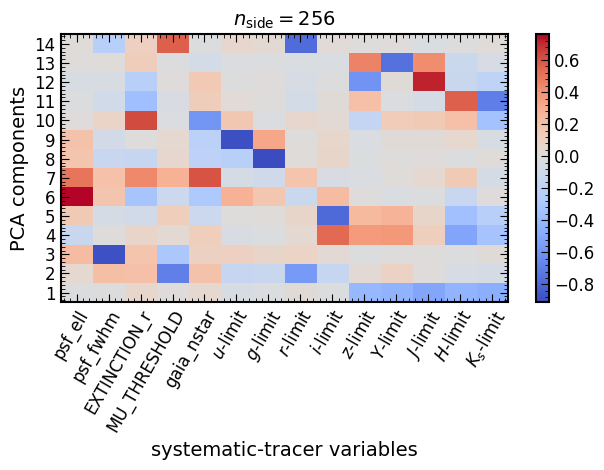}
    \caption[KiDS: matrix of linear coefficients mapping KiDS bright sample systematic-tracer variables onto their corresponding orthogonal principal components]{Linear coefficients as indicated by the colour bar, describing the transformation of systematic-tracer variable maps into principle component maps for $n_{\rm{side}}=256$.}
    \label{kids:fig:PCAcoeffs}
\end{figure}

After creating maps of each of the systematic-tracers listed in Table \ref{kids:tab:housekeeping_variables}, we masked pixels in the $1st^{\rm{}}$ and $99th^{\rm{}}$ percentile tails of each distribution and made additional cuts by hand, excluding the few sparse pixels with extreme outlier values for systematic-tracers. These would otherwise contaminate correlations with poorly sampled unrepresentative galaxy counts. We then whitened the parameter distributions (transforming each to zero-mean and unit variance) before retrieving 14 orthogonal principle component maps. The matrix of linear coefficients connecting PCA components to systematics tracers is shown in Fig. \ref{kids:fig:PCAcoeffs}.

\begin{figure*}
    \centering
    \includegraphics[width=\textwidth]{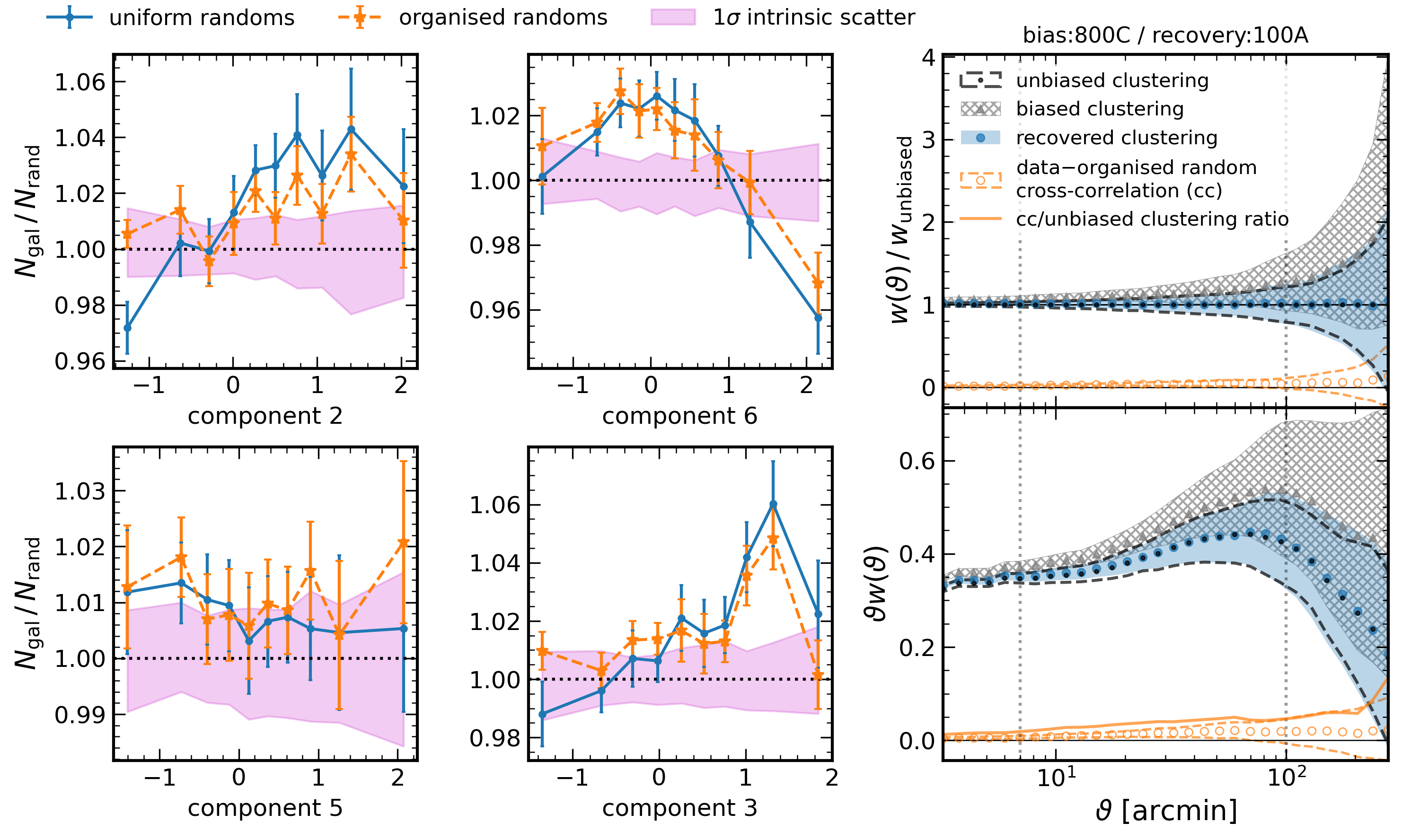}
    \caption{\hlstop{Illustration of divergent one- and two-point correlation corrections from organised randoms.} \emph{Left:} Pixel (one-point) correlations between the galaxy number density $N_{\rm{gal}}/N_{\rm{rand}}$ in biased (according to the \ttt{800C} SOM) \flask mocks and the mean PCA components in the same pixels (see Appendix \ref{kids:app:1point_correlations}). Correlations normalised by uniform randoms' counts-in-pixels are given in blue, and those normalised (corrected) by the pixel counts of \ttt{100A} organised randoms are given in orange. Correlations and errors are the mean and the root-diagonal of the covariance over \nflask \flask realisations, respectively. Purple shading indicates the expected intrinsic $1\sigma$ spread in these correlations, as calculated over 100 unbiased \flask realisations (Fig. \ref{kids:fig:expected_scatter_t3}). Shown here are the 4 (out of 14) correlations for which the organised randoms correction most improves consistency with unity. \emph{Right:} Corresponding biased and recovered two-point angular clustering correlations $w(\vartheta)$, shown in ratio to (\emph{top}) and overlaid on (\emph{bottom}) the unbiased clustering signature, as in Fig. \ref{kids:fig:t3_master}. We also show the cross-correlation between organised randoms and the unbiased data and its ratio to the unbiased clustering signal in orange. Correlations and errors are again the mean and the root-diagonal of the covariance over \nflask \flask realisations.
}
    \label{kids:fig:t2t3_800C100A}
\end{figure*}

\begin{figure*}
    \centering
    \includegraphics[width=\textwidth]{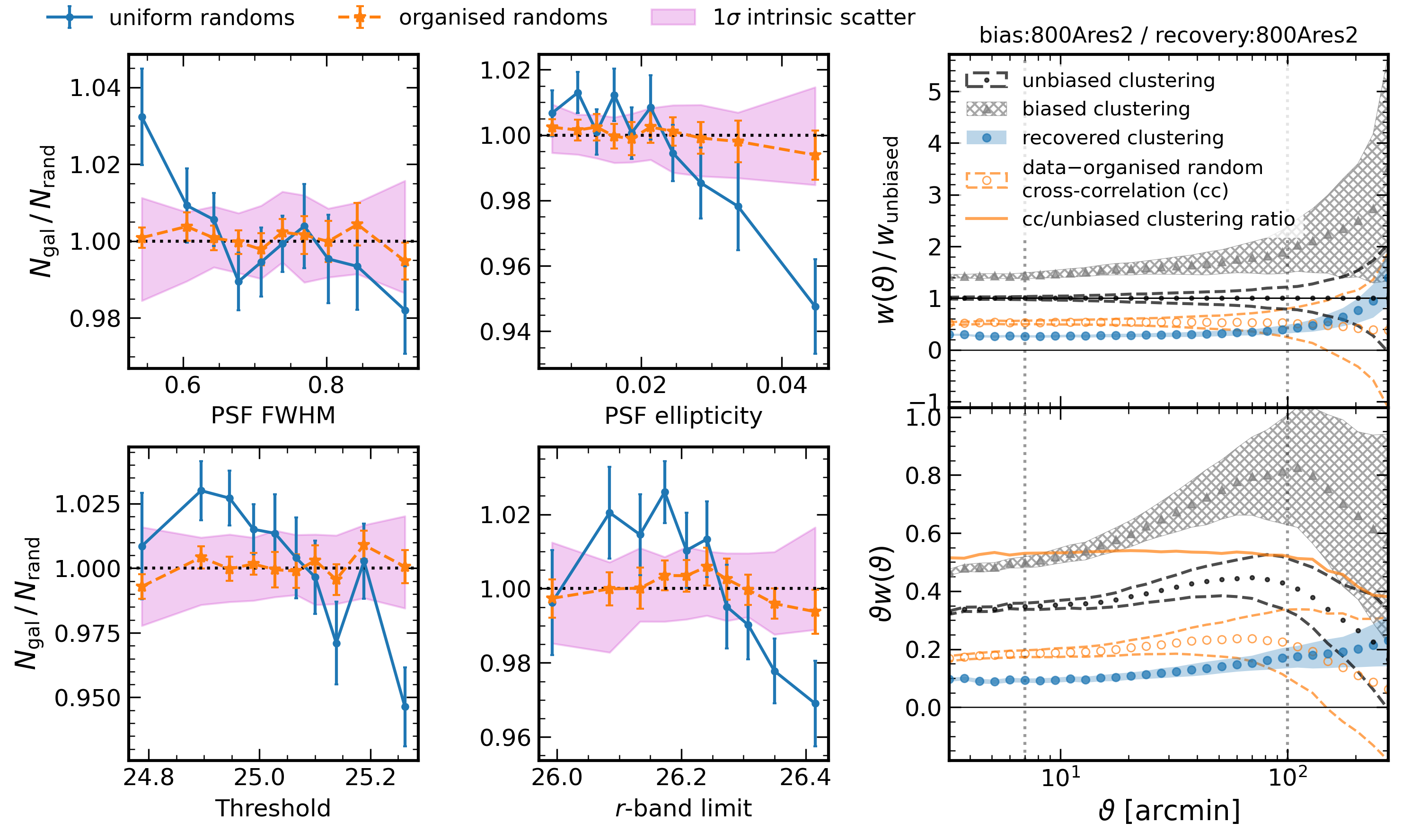}
    \caption{Same as Fig. \ref{kids:fig:t2t3_800C100A}, now for bias:\ttt{800Ares2} / recovery:\ttt{800Ares2,} \ie the high-resolution setup, and considering systematic-tracers rather than PCA components. Nearly perfect one-point correlation corrections (\emph{left}) are accompanied by the cross-correlation between unbiased \flask data and organised randoms (\emph{right;} orange) rising to $\sim50\%$ of the unbiased clustering signature, compared with the negligible cross-correlation in Fig. \ref{kids:fig:t2t3_800C100A}.
}
    \label{kids:fig:t2t3_800A800Afail}
\end{figure*}

\tbf{
Figs. \ref{kids:fig:t2t3_800C100A} and  \ref{kids:fig:t2t3_800A800Afail} illustrate why these one-point correlation metrics are not suitable for assessing the performance of our corrective randoms. In Fig. \ref{kids:fig:t2t3_800C100A} we select the subset of four component-density pixel correlations in bias:\ttt{800C} \flask realisations that are best corrected to unity after normalisation by the \ttt{100A} organised randoms' density in these pixels (Fig. \ref{kids:fig:PCAcoeffs} shows that these components are closely related to the PSF or threshold training variables from parameter set \ttt{A}). The implication is that residual correlations will cause two-point clustering correlations to be biased, as the \ttt{100A} SOM is insufficiently tracing systematic density fluctuations at the pixel level. However, the right-hand side in Fig. \ref{kids:fig:t2t3_800C100A} also shows that these organised randoms are capable of recovering unbiased clustering two-point functions.
}

\tbf{
Conversely, when we raise the grid resolution of our organised randoms (setup \ttt{800Ares2} from Table \ref{kids:tab:SOMs}), we see that the systematic density one-point correlations in Fig. \ref{kids:fig:t2t3_800A800Afail} are almost perfectly corrected (we show systematic-tracers in this figure as they are easier to interpret and because the component corrections are similarly nearly perfect), whilst the two-point functions reveal a pathological over-correction of the clustering signature. This is because the \ttt{800Ares2} randoms reproduce the real cosmic structure in the data too closely (demonstrated by the unbiased data-organised randoms cross-correlation, which is at $\sim50\%$ of the unbiased signal), and thus act to destroy real clustering signals. We therefore restricted our validation of organised randoms to two-point correlation statistics, which offer insight into potential over-corrections as organised randoms become contaminated by cosmic structure.
}

\section{Data-driven systematics}
\label{kids:app:datadriven_syst}

\begin{figure*}[!htp]
    \centering
        \includegraphics[width=\linewidth]{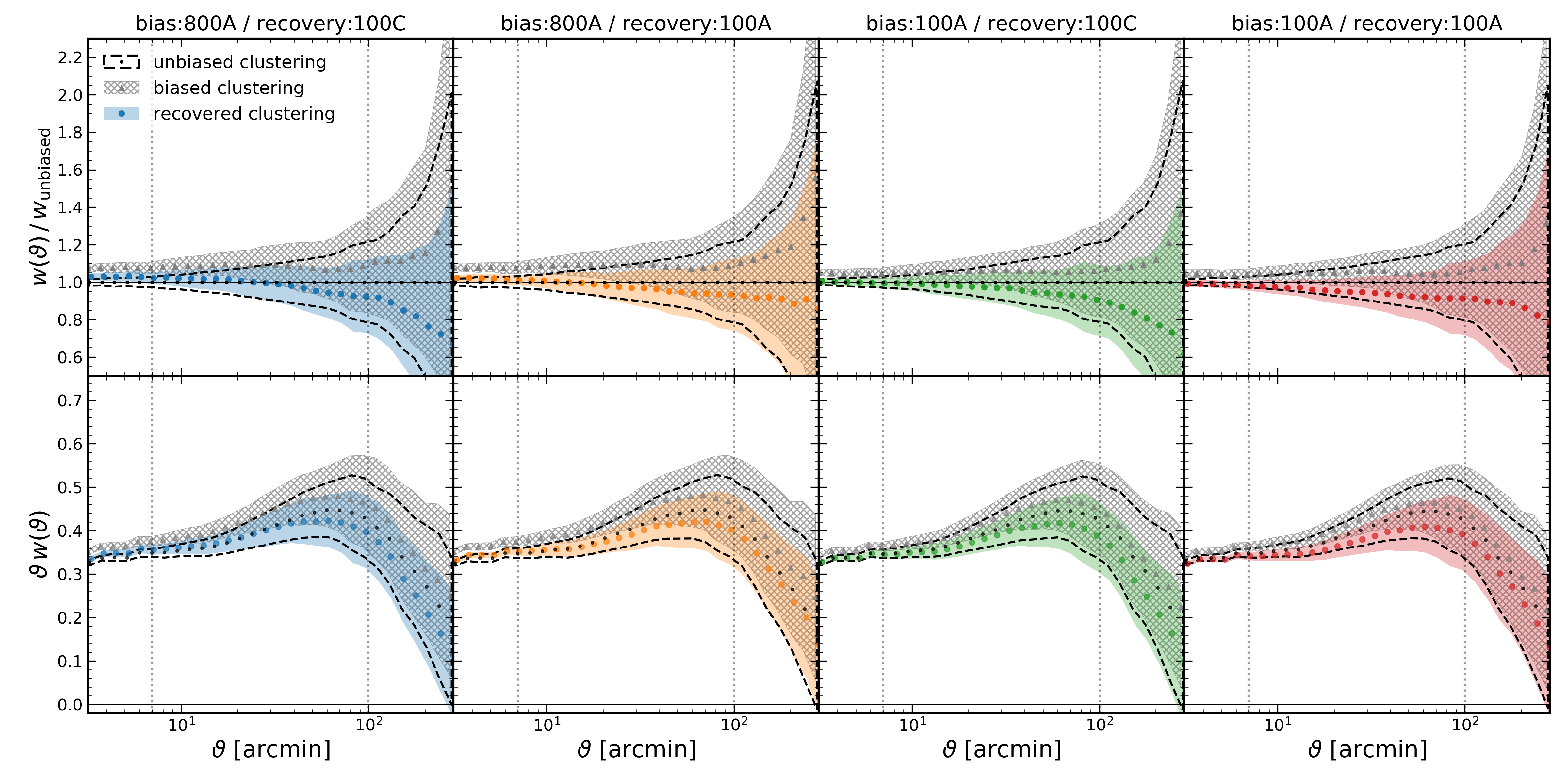}
    \caption[{\sc{flask}}: \wth signals in biased \flask fields, with/out corrective randoms -- additional configurations]{Same as Fig. \ref{kids:fig:t3_master}, now for some additional bias:SOM and recovery:SOM configurations, as indicated by panel titles (with reference to Table \ref{kids:tab:SOMs}).
    }
    \label{kids:fig:t3_extra}
\end{figure*}

\tbf{
Fig. \ref{kids:fig:t3_extra} presents additional configurations of the bias:[\ttt{800A,100A}] / recovery:[\ttt{100A,100C}] two-point correlations. As stated in the main text, each meaningful case yields a recovered clustering signal that is more consistent with the unbiased clustering. For bias:\ttt{800A} cases, where the bias is slightly weaker than for bias:\ttt{800C} (Fig. \ref{kids:fig:t3_master}) due to the omission of \ttt{MAG\_LIM\_r,EXTINCTION\_r,gaia\_nstar} parameters from training, the performance of organised randoms is similarly strong, and a slight preference for recovery:\ttt{100A} over \ttt{100C} is negligible with respect to the noise level ($\sim0.1\sigma$). When clustering biases are particularly weak, as for the bias:\ttt{100A} case, our methods become more prone to over-correction. Still, the mean absolute deviation from the unbiased case (in units of the standard deviation of the unbiased clustering; summarised for all configurations in Table \ref{kids:tab:test3_summary}) is improved in the recovery with respect to the biased case here, primarily due to the success of corrections at lower values of $\vartheta$. We note that the bias:\ttt{100A} case constitutes our most optimistic bias scenario, having low sensitivity to small-$\vartheta$ (via small $N_{\rm{HC}}$) and the minimum set of systematic-tracer training variables (\ttt{psf\_fwhm,psf\_ell,MU\_THRESHOLD}). For more pessimistic scenarios, the organised randoms' performance is consistently improved. As also stated in the text, this bodes extremely well for the potential salvage of unbiased clustering signals from heavily systematics-contaminated data, such as the faint KiDS-$1000$ shear sample \tr{(see Wright et al., in prep.)}.
}

\begin{table}[!h]
    \centering
    \caption{Relative performance of corrections to angular clustering correlations \wth on the intermediate scales $7<\vartheta<100\,\rm{arcmin}$ for differently configured clustering biases and organised randoms.}
    \begin{tabular}{cccc}
    \hline
        bias:SOM & recovery:SOM & biased & recovered \\
    \hline
    \hline
\ttt{100A} & \ttt{100A} & $0.69\sigma$ & $0.61\sigma$ \\
\ttt{100A} & \ttt{100C} & $0.79\sigma$ & $0.35\sigma$ \\
\ttt{100C} & \ttt{100A} & $0.94\sigma$ & $0.28\sigma$ \\
\ttt{100C} & \ttt{100C} & $0.96\sigma$ & $0.54\sigma$ \\
\ttt{100C(x4)} & \ttt{100C} & $12.17\sigma$ & $0.34\sigma$ \\
\ttt{800A} & \ttt{100A} & $1.13\sigma$ & $0.27\sigma$ \\
\ttt{800A} & \ttt{100C} & $1.19\sigma$ & $0.31\sigma$ \\
\ttt{800Ares2} & \ttt{800Ares2} & $7.62\sigma$ & $9.58\sigma$ \\
\ttt{800C} & \ttt{100A} & $1.43\sigma$ & $0.09\sigma$ \\
\ttt{800C} & \ttt{100C} & $1.48\sigma$ & $0.31\sigma$ \\


        
    \hline
    \end{tabular}
    \tablefoot{With reference to Table \ref{kids:tab:SOMs}, the bias:SOM column gives the SOM trained against KiDS-Bright to infer $\delta_{\rm{OR}}$ (Eq. \ref{kids:eq:Pdepletion}) and inform the density-field bias applied to \nflask \flask realisations of KiDS-Bright-like data. The recovery:SOM column gives the SOM trained against those biased \flask mocks to produce organised randoms (see Sec. \ref{kids:sec:flask_biasing} for details). On average over the \nflask \flask realisations, the biased and recovered columns then give the absolute deviation from the unbiased clustering signal $|\,w_{\rm{biased/recovered}}-w_{\rm{unbiased}}\,|$ in units of the uncertainty in the unbiased signals $\sigma$, averaged over the intermediate scales $7-100\,\rm{arcmin}$. Non-uniformity of biased column values with self-similar bias:SOM setups arises stochastically due to our probabilistic application of biases to \flask realisations, and all variations are at $<0.1\sigma$. The \ttt{800Ares2} configuration illustrates a failure mode of our organised randoms methods, where the resolution and scale sensitivity of the setup are too high, resulting in over-fitting to cosmic structure and a consequent over-correction of the clustering signal. 
    }
    \label{kids:tab:test3_summary}
\end{table}

\section{GAMA comparisons}
\label{kids:app:gama_comparison}

\begin{figure}
    \centering
    \includegraphics[width=\columnwidth]{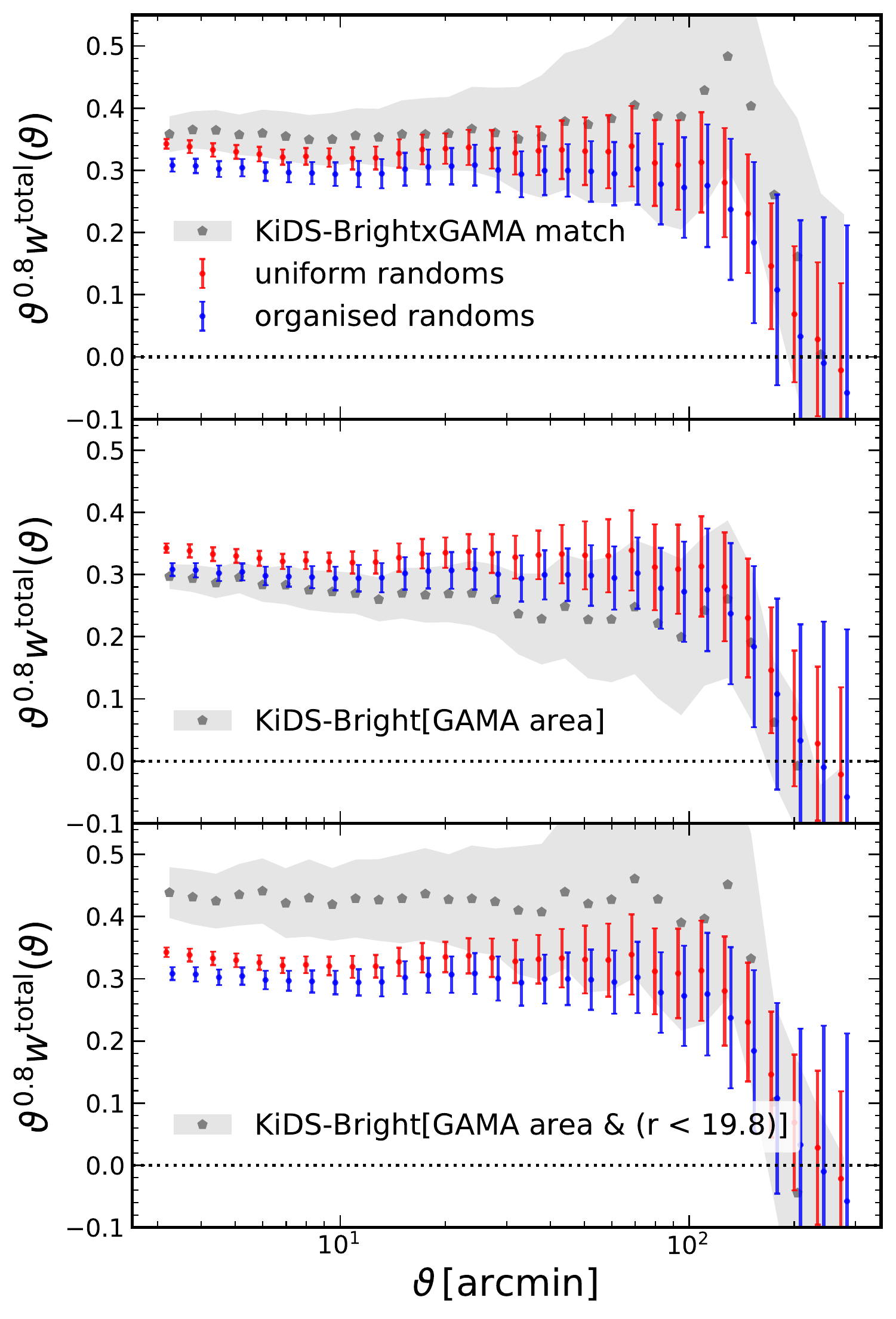}
    \caption{Comparison of different clustering measurements in the GAMA area and how they compare with the KiDS-Bright correlations measured with (blue) and without (red) \ttt{100A} organised randoms. Red and blue data-points are the same in all rows, whilst grey points and shading are the angular clustering measured with uniform randoms for (\emph{top}) the KiDS-Bright-GAMA cross-matched catalogue, (\emph{middle}) KiDS-Bright galaxies within the GAMA window, and (\emph{bottom}) KiDS-Bright galaxies within the GAMA window, and with $r<19.8$. Errors are estimated with delete-one jackknifes from 31 sub-regions of KiDS-Bright or 20 sub-regions of the GAMA window. Errors on scales $\vartheta\gtrsim180\,\rm{arcmin}$ (the approximate scale of each sub-region) are thus likely to be underestimated for the GAMA-area correlations.
    }
    \label{kids:fig:gama_comparison}
\end{figure}

As discussed in Sec. \ref{kids:sec:kids}, our KiDS-Bright selection is deliberately GAMA-like for the training of photo-$z$ estimation \citep{Bilicki2021}. Thus we might expect a similar clustering signature from KiDS-Bright, but for a reduction in amplitude due to photometric redshift scatter. Scattering disconnected and connected galaxies in and out of each bin will dilute the measurable clustering signals. To replicate this effect in the spectroscopic GAMA sample, we cross-matched GAMA with KiDS-Bright and defined redshift bins using the KiDS {\sc{ANNz2}} photo-$z$. The resulting total correlations, shown as grey points and shading in the top-panel of Fig. \ref{kids:fig:gama_comparison}, exceeded even the uncorrected KiDS-Bright correlations in amplitude. We determined that KiDS-Bright is not sufficiently GAMA-like to expect agreement here by remeasuring correlations for \emph{\textup{all}} KiDS-Bright galaxies in the GAMA window (Fig. \ref{kids:fig:gama_comparison}; middle panel), and finding a reduced clustering amplitude which is more consistent with the corrected (by organised randoms) correlation. We measured correlations once more for KiDS-Bright in the GAMA window, now with a flux limit $r<19.8$ applied to KiDS magnitudes\footnote{KiDS magnitudes are not equivalent to the Petrosian magnitudes against which GAMA targets were selected, therefore this flux limit is only approximately GAMA-like \citep[see][]{Bilicki2018}.} (Fig. \ref{kids:fig:gama_comparison}; bottom panel), and again see the clustering amplitude rise to exceed all KiDS-Bright signals. We conclude that a comparison with the GAMA auto-clustering signals is not entirely valid for KiDS-Bright, noting also that with only $\sim1/5$ of the KiDS-Bright area, the impacts of increased sample variance in GAMA present further complications.

\end{document}